\documentclass[preprint]{aastex63}

\usepackage{amsmath}
\usepackage{xcolor}
\definecolor{darkgreen}{RGB}{0,127,0}

\newcommand{\grad}{\boldsymbol{\nabla}}
\newcommand{\cross}{\boldsymbol{\times}}

\providecommand{\e}[1]{\ensuremath{\times 10^{#1}}}

\received{\today}
\revised{\today}
\accepted{\today}
\submitjournal{ApJ}

\shorttitle{Correlated Evolution of Flare EUV Ribbons and HXR Emissions}
\shortauthors{}
\graphicspath{{./}{figures/}}

\begin{document}

\title{Correlated Spatio-temporal Evolution of Extreme-Ultraviolet Ribbons and Hard X-rays in a Solar Flare}
\author{Stephen J.\ Naus}
\affiliation{Department of Physics, University of Maryland, College Park, MD 20742, USA}
\affiliation{Heliophysics Science Division, NASA Goddard Space Flight Center, Greenbelt, MD 20771, USA}
\author{Jiong Qiu}
\affiliation{Department of Physics, Montana State University, Bozeman, MT 59717, USA}
\author{C. Richard DeVore}
\affiliation{Heliophysics Science Division, NASA Goddard Space Flight Center, Greenbelt, MD 20771, USA}
\author{Spiro K. Antiochos}
\affiliation{Heliophysics Science Division, NASA Goddard Space Flight Center, Greenbelt, MD 20771, USA}
\author{Joel T. Dahlin}
\affiliation{Heliophysics Science Division, NASA Goddard Space Flight Center, Greenbelt, MD 20771, USA}
\author{James F. Drake}
\affiliation{Institute for Research in Electronics and Applied Physics, University of Maryland, College Park, MD 20742, USA}
\author{M. Swisdak}
\affiliation{Institute for Research in Electronics and Applied Physics, University of Maryland, College Park, MD 20742, USA}

\begin{abstract}

We analyze the structure and evolution of ribbons from the M7.3 SOL2014-04-18T13 flare using ultraviolet (UV) images from
\emph{IRIS} and \emph{SDO}/AIA, magnetic data from \emph{SDO}/HMI,
hard X-ray (HXR) images from \emph{RHESSI}, and light curves from
\emph{Fermi}/GBM, in order to infer properties of coronal magnetic reconnection.  As the
event progresses, two flare ribbons spread away from the magnetic
polarity inversion line.  The width of the newly brightened front
along the extension of the ribbon is highly intermittent in both space
and time, presumably reflecting non-uniformities in the structure
and/or dynamics of the flare current sheet. Furthermore, the ribbon
width grows most rapidly in regions exhibiting concentrated
non-thermal HXR emission, with sharp increases slightly preceding the
HXR bursts.  The light curve of the ultraviolet emission matches the
HXR light curve at photon energies above 25 keV.  In other regions the
ribbon-width evolution and light curves do not temporally correlate with the HXR emission.  
This indicates that the production of non-thermal
electrons is highly nonuniform within the flare current sheet.  Our
results suggest a strong connection between the production of
non-thermal electrons and the locally enhanced perpendicular extent of
flare ribbon fronts, which in turn reflects inhomogeneous structure
and/or reconnection dynamics of the current sheet. Despite this
variability, the ribbon fronts remain nearly continuous,
quasi-one-dimensional features.  Thus, although the reconnecting
coronal current sheets are highly structured, they remain
quasi-two-dimensional and the magnetic energy release occurs
systematically, rather than stochastically, through the volume of
reconnecting magnetic flux.  

\end{abstract}


\section{Introduction} 
\label{sec:intro}

Flare energy release is governed by fast magnetic reconnection taking place in the corona \citep{Priest2002}.
For a few decades, the standard model for eruptive two-ribbon flares, known as the Carmichael-Sturrock-Hirayama-Kopp-Pneuman (CSHKP) model 
\citep{Carmichael1964, Sturrock1966, Hirayama1974, Kopp1976}, has been established and
supported by numerous observations. A prominent signature of these flares is the formation of an arcade of flare loops growing in the corona and a pair of flare ribbons in the chromosphere spreading apart and away from the magnetic polarity inversion line (PIL). According to the standard model, the separating two ribbons outline the feet of the growing arcades formed by reconnection as it
proceeds along a vertical current sheet trailing an erupting flux rope. The direction of the current in the current sheet is presumed to follow along the PIL or the extension of the two ribbons, and the leading edges of the moving ribbons map the feet of two sets of magnetic field lines of opposite signs that are reconnecting \citep{Svestka1980, Forbes1984}. The model also
schematically describes the magnetic configuration and temporal evolution of energized particles
and plasmas in the corona, as well as the dynamics of the lower atmosphere in response to energy deposition \citep[see Figure 1 in][]{Forbes1996}. The properties of these particles, plasmas, and the atmospheric dynamics have been diagnosed using observations in many different wavelengths including radio, optical, ultraviolet (UV), extreme ultraviolet (EUV),
soft X-ray (SXR), and hard X-ray (HXR) \citep[e.g., see the review by][]{Fletcher2011}. Recently, an extraordinary eruptive flare, SOL2017-09-10 X8.2, exhibited a wide range of observational
signatures broadly consistent with the standard model, such as the erupting flux rope \citep{Seaton2018, Long2018, Veronig2018, Gopalswamy2018}, the trailing current sheet \citep{Seaton2018, Warren2018, Longcope2018}, energetic electrons produced at the flare loop top that deposit energy all along the flare loop and produce bremsstrahlung HXR emissions and gyro-synchrotron microwave emissions \citep{Gary2018,Chen2020}, and energetic particles accelerated by the eruption causing ground level enhancement events (GLE) \citep{Guo2018, Gopalswamy2018}. In particular, advanced microwave observations with the Expanded Owens Valley Solar Array \citep[EOVSA;][]{Gary2013} have yielded crucial measurements of plasma and magnetic field properties around the current sheet where flare
reconnection takes place \citep{Chen2020}. These include the magnetic field strength and its evolution, plasma inflows, and the reconnection rate in terms of the reconnection electric field and the inflow Alfv\'en Mach number. The observations provide strong support to the standard model. 

Despite its success, the standard model is a global two-dimensional model \citep[also see its three-dimensional (3D) variant in][] {Aulanier2012}. Magnetic reconnection in solar flares is inherently 3D, and even flares with two coherent
ribbons have demonstrated many features indicative of the 3D nature of flare reconnection.
The supra-arcade downflows \citep[SADs;][]{McKenzie1999, Savage2011} discovered in limb observations
clearly depart from an organized laminar motion, and have been considered as evidence for patchy reconnection,
albeit in the context of the global reconnecting current sheet \citep[e.g.][]{Klimchuk1996}.
Viewed from above and against the disk, a flare arcade is not monolithic but is a collection of discrete
loops of finite cross-sections, as demonstrated by high-resolution EUV imaging observations \citep[e.g.][]{Aschwanden2001}.
Similarly, flare ribbons in the chromosphere, typically observed in optical, UV, or EUV images, consist
of patches or kernels of various sizes and brightness that exhibit temporal and spatial
evolution deviating from the simple picture of global laminar structure and evolution of flare ribbons \citep[e.g.][]{Warren2001,
Fletcher2004, Brannon2015, Graham2015, Jing2016}.

Furthermore, whereas optical and UV observations often show two or more extended ribbons,
 two-ribbon hard X-ray flares have rarely been observed 
\citep{Masuda2001, Liu2007b, Krucker2011}. Images of thick-target bremsstrahlung hard X-ray emissions
by non-thermal electrons typically reveal several kernels \citep[e.g.][]{Sakao1994, Bogachev2005, Yang2009}. It is likely that the present instruments, which have limited dynamic range, cannot measure weak hard X-ray sources due to non-thermal electrons even if they are ubiquitous \citep[e.g.][]{Testa2014, Glesener2020}. It is also likely that not all optical or UV brightenings in the lower atmosphere are produced by non-thermal electrons. Other energy transport mechanisms, such as thermal conduction \citep{Gan1991, Longcope2014} and Alfv\'{e}n waves \citep{Fletcher2008, Kerr2016, Reep2018}, may carry energy from the corona, where it is released by reconnection, to the feet of reconnection-formed loops. Either way, these observations suggest that reconnection and subsequent energy transport are not uniform along the reconnecting current sheet.

Theoretical and computational models of particle energization during reconnection suggest that strong particle energy gain is a consequence of the development of multiple flux ropes in reconnecting current sheets \citep{drake06b,drake13a,dahlin15a,dahlin17a,li19b}. That reconnecting current sheets develop complex, multi-island structures has been established both in MHD  \citep{biskamp86a,daughton09a,bhattacharjee09a,cassak09a,huang10a} and kinetic \citep{drake06b,daughton11a} simulations. The structuring of current layers through the formation of flux ropes has also been documented in high Lundquist-number global MHD simulations \citep{karpen12a,guidoni16a,dahlin21a}. Finally, the recent development of new computational models for exploring reconnection-driven particle energy gain in macroscopic systems has revealed extended power-law distributions of electrons driven by the fragmentation of the reconnecting current layer through flux-rope formation and merging \citep{arnold21a}. Thus, establishing observational evidence for the structuring of coronal current sheets and the connection between such structures and hard X-ray production is essential to bring closure to the linkage between reconnection-driven magnetic energy release and particle energization.

To study the three-dimensional properties of magnetic reconnection, 
we analyze imaging observations of a two-ribbon flare, SOL2014-04-18T13:03, obtained by the {\em Interface Region Imaging Spectrograph} \citep[{\em IRIS};][]{DePontieu2014}. By tracking the evolution of flare ribbons in the chromosphere, we infer properties of magnetic reconnection in the corona, and explore the relationship between these properties and the UV and HXR emission. In the following text, we provide an observational overview of the flare in
multiple wavelengths (\S\ref{sec:overview}). We then analyze newly brightened flare ribbon fronts derived from the time sequence of the {\em IRIS} slit-jaw images (SJIs), study the spatio-temporal evolution of these features, and compare them with the UV and HXR emissions (\S\ref{sec:analysis}). We discuss and interpret our results in \S\ref{sec:discussion}, and conclude the paper by summarizing our principal findings in \S\ref{sec:conclusions}.

\section{Overview of Observations} 
\label{sec:overview}


The eruptive two-ribbon flare SOL2014-04-18T13:03 was observed by {\em IRIS}, the {\rm Extreme-ultraviolet Imaging Spectrometer} \citep[EIS;][]{Culhane2007} on {\em Hinode}, the {\it Reuven Ramaty High Energy Solar Spectroscopic Imager} \citep[{\em RHESSI}; ][]{Lin2002}, the {\rm Gamma-ray Burst Monitor} \citep[{\rm GBM};][]{Meegan2009} on {\em Fermi}, and the {\rm Atmospheric Imaging Assembly} \citep[{\rm AIA};][]{Lemen2012} on the {\em Solar Dynamics Observatory} \citep[{\em SDO};][]{Pesnell2012}. EIS and {\em IRIS} observations provide spectroscopic diagnostics of flare plasmas at certain locations across some parts of the flare ribbons. The spectroscopic observations were analyzed by \citet{Brannon2015}, \citet{Brosius2015}, \citet{Brosius2016}, and \citet{Mulay2021}, revealing quasi-periodic pulsations in the time profiles of the intensity and Doppler-shifted upflows and downflows of the plasma.

Imaging observations by AIA and {\em IRIS} capture the temporal and spatial evolution of the flare in multiple passbands. In particular, the structure and evolution of flare ribbons in the lower atmosphere indicate the progress of magnetic reconnection and energy release in the corona. This study focuses on the analysis of flare ribbons using the {\em IRIS} SJIs in the far ultraviolet (FUV) band at 1400~\AA\ and near ultraviolet (NUV) band at 2796~\AA. Flare emission in the FUV image is dominated by the Si {\sc iv} line formed in the transition region at the characteristic temperature $T_{\rm 1400} \approx 63,000$~K, and in the NUV images by the Mg {\sc ii} k line formed in the upper chromosphere ($T_{\rm 2796} \approx 10,000$~K). These images were obtained with the pixel scale of 0.17\arcsec\ and a moderate time cadence of 28~s. 
In addition, the flare produced hard X-ray emissions at photon energies up to 100~keV, which were observed by {\em RHESSI} and {\em Fermi}/GBM.

\begin{figure}[ht!]
\plotone{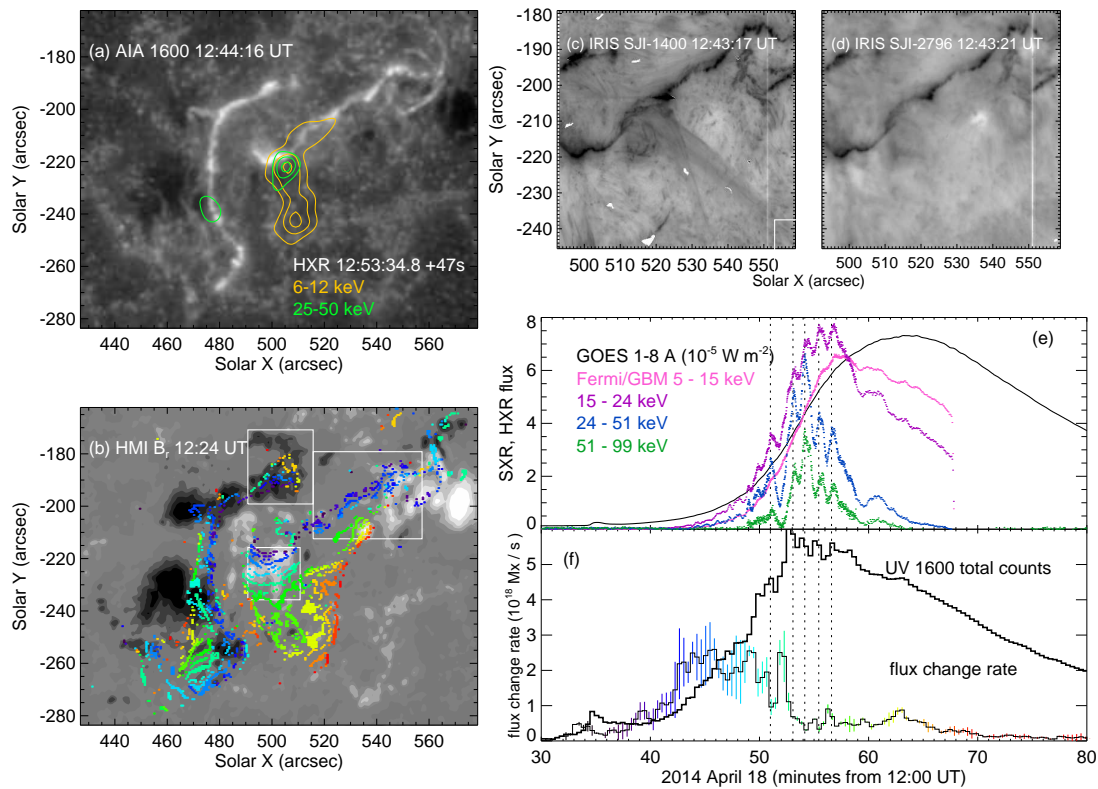}
\caption{Overview of the flare. (a) A snapshot of the flare ribbons observed in AIA 1600~\AA\ passband during the rise of the flare; superimposed are \emph{RHESSI} HXR maps at 6-12 keV (orange) and 25-50 keV (green) at the peak time of the flare, using contour levels at 0.4, 0.65, and 0.9 of maximum intensity in each map.
(b) Evolution of newly brightened ribbon fronts, at every two minutes, mapped in a magnetogram of the radial component of the vector magnetic field by HMI. The background grey scales represent magnetic flux densities of $\pm 200, 400, 800, 1200, 1600$ Mx~cm$^{-2}$. The color of the ribbon symbols indicates the time of the ribbon fronts as shown in panel (f). The three rectangular boxes denote the regions in which flare ribbon fronts are analyzed using {\em IRIS} SJI images (see Section 2.2). (c)-(d) Snapshots of {\em IRIS} SJI in FUV (1400~\AA) and NUV (2796~\AA), respectively. {The white box in (c) indicates the position of a reference quiescent region. The flare ribbon brightness is normalized to the median brightness of this region. See \S\ref{subsec:fronts}.} 
(e-f) The total light curves of the flare in SXR {(in units of 10$^{-5}$ W m$^{-2}$)}, HXR, and UV 1600~\AA, as well as the total flux change rate $\dot{\psi}$ derived from the AIA 1600~\AA\ images and the HMI magnetogram illustrated in (b). {The HXR and UV light curves are arbitrarily normalized.}   
The heliographic coordinates shown in panels (a-d) represent the position of the flare region at 12:40:16~UT. 
\label{fig:overview}}
\end{figure}

Figure~\ref{fig:overview}a displays a snapshot of the two flare ribbons observed by AIA in the 1600~\AA\ passband. Flare emission in this passband includes the enhanced emission of the optically thin C {\sc iv} line formed at the transition region temperature ($T_{\rm 1600} \approx 100,000$~K), as well as other chromosphere lines and continuum \citep{Simoes2019}.
The newly brightened flare ribbon fronts are derived from the time series of the 1600~\AA\ images \citep[e.g.][]{Saba2006, Qiu2010} and are displayed in Figure~\ref{fig:overview}b, superimposed on a magnetogram of the radial component $B_r$ of the magnetic field measured by {\em SDO}'s Helioseismic and Magnetic Imager \citep[HMI; ][]{Schou2012}. 
In the figure, symbols from cool to warm colors denote newly brightened ribbon fronts during the flare evolution from 12:30 to 13:10~UT. More details on how the ribbon fronts are calculated are presented in \S\ref{subsec:fronts}. The morphological evolution is suggestive of the global organization of two ribbons that are extended along the PIL; on the other hand, different locations along the two ribbons do not brighten uniformly. The fine-scale ribbon structure will be explored using  high-resolution SJIs, shown in Figure~\ref{fig:overview}c,d.\footnote{The coalignment between the {\em IRIS} SJIs and SDO/AIA or HMI was made by the IRIS team. A cross-correlation between the AIA-1600 and {\em IRIS} SJIs in different areas in the field of the view of the SJIs suggests that this automated coalignment is as good as 0.8\arcsec; therefore, we adopt the automated coalignment from the IRIS software, and do not adjust the coalignment further.} Note that to better show the details of the flare ribbons, these two panels only present a partial frame of the {\em IRIS} SJIs in the rise phase of the flare. The full field of view of {\em IRIS} SJI covers the entire ribbon in the regions of positive line-of-sight photospheric magnetic field throughout the flare evolution. 
The fine-scale structures of flare ribbons reveals non-uniform reconnection and energy release in the corona and will be explored in the following analysis.

The overall evolution of the flare energy release is illustrated in the total light curves in various wavelengths. Figure~\ref{fig:overview}e shows the UV, HXR (15-100 keV), and SXR (1-8~\AA) light curves of the flare, obtained by AIA in the 1600~\AA\ passband, {\em Fermi}/GBM, and the {\em Geostationary Operational Environmental Satellites} ({\em GOES}), respectively. Significant HXR emissions, produced by non-thermal electrons impacting the lower atmosphere, occur between 12:50 and 13:00~UT, and exhibit multiple bursts with quasi-periodicities between 1 and 2 minutes. Such quasi-periodic pulsations were also observed and studied in EIS, {\em IRIS}, and radio spectroscopic observations \citep{Brannon2015, Brosius2016, Karlicky2017, Mulay2021}. Five of these bursts are indicated in the figure. 

Flare energy release is believed to be governed by magnetic reconnection in the corona. The amount of magnetic flux undergoing reconnection per unit time, or the flux change rate $\dot{\psi}$, can be measured by summing up the magnetic flux swept up by newly brightened ribbons in the lower atmosphere \citep{Forbes2000, Hesse2005}. This is shown in the lower panel of Figure~\ref{fig:overview}f, with the same color code for time as used in Figure~\ref{fig:overview}b. Uncertainties in the $\dot{\psi}$ plot are obtained using varying thresholds of flare ribbon fronts from 4 to 6 times the quiescent background intensity; $\dot{\psi}$ is measured in both positive and negative magnetic fields \citep{Qiu2007, Qiu2010, Kazachenko2017}, and the average of these measurements is used.\footnote{In this calculation, the pixel area is not de-projected, nor is the magnetic field extrapolated to the chromosphere where the ribbons form, as the corrections from these two effects would partly cancel each other. Uncertainties due to these effects are discussed 
by \citet[][]{Qiu2007}; see also references therein.} 

In this flare, $\dot{\psi}$ rises together with the UV emission, and peaks 7 minutes later, when the UV emission is still half way towards the peak, and the $>25$ keV HXR just starts to rise. At its peak, $\dot{\psi}$ is about 4-5 times greater than 10 minutes later, when HXR and UV emissions reach their maxima. This is different from some other studies showing a temporal correlation between $\dot{\psi}$ and non-thermal HXR or microwave light curves \citep[e.g.,][]{Qiu2004, Miklenic2009}. 
Furthermore, the non-thermal HXR emission appears to only involve localized regions along the flare ribbons, as indicated by the \emph{RHESSI} HXR maps shown in Figure~\ref{fig:overview}a \citep[also see][]{Brosius2016}. At its peak, the 25-50 keV HXR emission is concentrated in two kernels on the two flare ribbons in magnetic fields of opposite polarities while the HXR emission at 6-12 keV traces a larger part of the positive ribbon and is extended between the two ribbons. Such a configuration suggests that higher energy ($>$25 keV) HXRs are thick-target footpoint sources, whereas the emission at the lower energy ($<$20 keV) include contributions from the flare loop connecting the conjugate footpoints. 

These observations suggest that, in this two-ribbon flare, the amount of energy release in general, as reflected in the total UV emission, and the non-thermal energy release in particular, as indicated by the HXR emission, are not related in a simple way to the {\em global} properties of flare reconnection, such as the flux change rate. Significant non-thermal emission occurs more than 10 minutes after the start of the flare and only involves localized regions along the flare ribbons. In the following section, we will establish the
spatio-temporal structure of flare ribbons using high-resolution {\em IRIS} data and
examine its relationship with the flare energetics.




\section{Analysis and Results}
\label{sec:analysis}
It is widely accepted that flare ribbons map the locations of reconnection-associated energy release in the corona down to the lower transition region and chromosphere. Consequently, detailed analyses of the time sequence of flare ribbon observations potentially provide diagnostics of the progress of coronal reconnection. In this study, we focus on the temporal and spatial evolution of the {\em leading edges} of the flare ribbons, or the {\em ribbon fronts}, which we consider to be the surface signatures of newly reconnected magnetic field lines in the coronal flare current sheet \citep{Svestka1980, Forbes1996, Hesse2005}. These measurements can help probe the evolution and structure of the coronal reconnecting current sheet. For this event, the cadence of the {\em IRIS} SJIs is 28~s, which determines the minimum time resolution that can be achieved in this study.

\subsection{Determination of Flare Ribbon Fronts}
\label{subsec:fronts}
We define ribbon fronts by the first time that a pixel exceeds a brightness threshold. Specifically, a pixel is considered to be activated when its brightness $I$ reaches $N$ times the median brightness, $I_b$, of the quiescent background\footnote{During the {\em IRIS} observation period between 12:33 and 14:07~UT, the SJI-2796 were taken with a constant exposure $\tau = 8.0$~s, whereas the SJI-1400 were taken with a varying exposure $\tau = 0.4 - 8.0$~s. Therefore, in the analysis, we keep the brightness of the SJI-2796 images in units of counts, or data numbers (DNs), whereas the brightness of the SJI-1400 images is normalized to the exposure to be in units of DN s$^{-1}$. The quiescent background brightness $I_b$ is the median brightness of a non-flaring region of size 42\arcsec\ by 33\arcsec\ centered at the heliographic position (574\arcsec, -254\arcsec). Between 12:33 and 14:07~UT, the median $I_b$ in this quiescent region is 144$\pm$2 DN for SJI-2976, and 6$\pm$1 DN s$^{-1}$ for SJI-1400. In comparison, the median brightness of the AIA-1600 in this same quiescent region during the same time period is 144$\pm$3 DN at a constant exposure $\tau = 2.9$~s.} and remains bright for at least 4 minutes. The first criterion distinguishes flaring pixels from non-flaring features, such as active region plages \citep[e.g.][]{Saba2006, Qiu2010}, and the second selects pixels at the feet of {\rm closed} flare loops whose brightness decays gradually, typically over at least several minutes \citep{Cheng2012, Liu2013, Qiu2016, Graham2020}. This approach minimizes inclusion from other features, such as ejecta or pixels contaminated by saturation which brighten only briefly. The threshold scaling factor, $N$, is determined empirically from the histogram of brightness of the flaring region throughout the evolution. {With these criteria, flaring pixels are picked out in each image and a mask is generated marking these pixels, and newly brightened ribbon fronts are obtained from the difference of two consecutive masks.}

\begin{figure}[ht!]
\plotone{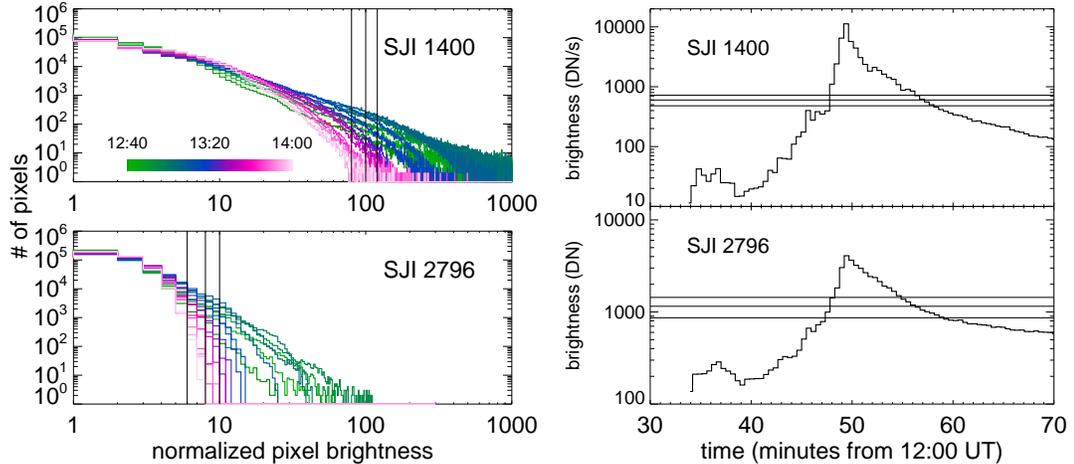}
\caption{Left: histograms of pixel brightness, normalized to the brightness of the quiescent region, measured in SJI-1400 (top) and SJI-2976 (bottom) during the flare evolution from the early phase (cool colors) to the post-flare phase (red). The time color scale is indicated in the top panel. Vertical lines mark the threshold brightness at which the flare brightness sufficiently exceeds the post-flare (reference) brightness. Right: light curves of a selected flaring pixel in the two passbands, 1400~\AA\ and 2796~\AA\ from {\em IRIS}-SJI. The horizontal lines indicates the threshold $N = 80, 100, 120$ for SJI-1400, and the blue lines indicate the thresholds $N = 6, 8, 10$ for SJI-2796. These thresholds are used to determine the ribbon fronts; see text for details.
\label{fig:lgtcv}}
\end{figure}

The left panels in Figure~\ref{fig:lgtcv} show the temporal development of the brightness histogram, normalized to $I_b$ at each time, for SJI-1400 and SJI-2796. The color indicates the time during the flare, from the early phase (cool colors) to the late phase (warm colors). For this event, {\em IRIS} observations started when the flare was already in progress; therefore, we use the post-flare histogram (red) as the reference for the non-flare brightness distribution. In the example shown in the figure, we distinguish the flare brightness by adopting $N$ = 100 for SJI-1400 and $N$ = 8 for SJI-2796 to set the threshold for identifying newly brightened ribbon pixels. To examine the sensitivity of our measurements to the threshold, we also extended the measurements using $N$ = 80 and 120 for SJI-1400 and $N$ = 6 and 10 for SJI-2796. Using varying thresholds helps to explore the uncertainties introduced by instrument effects, such as scattered light, saturation, or varying exposures, {as their effects on the measurements cannot be accurately determined.}

The right panels in Figure~\ref{fig:lgtcv} show the light curves of a single pixel in the {\em IRIS} SJI in both the NUV and FUV passbands. Bright flare-ribbon pixels typically exhibit a rapid rise in emission and reach their maximum within 2 minutes, or several time frames of the SJIs that were obtained at the cadence of 28~s. The thresholds $N$ = 6, 8, and 10 for SJI-2796 and $N$ = 80, 100, and 120 for SJI-1400 are marked in the figure to indicate this flaring pixel's time of activation, which in turn is assumed to be a measure of the time of reconnection energy release in the corona.

\subsection{Spatio-temporal Evolution of Ribbon Width}
\label{subsec:widths}
\begin{figure}[ht!]
\plotone{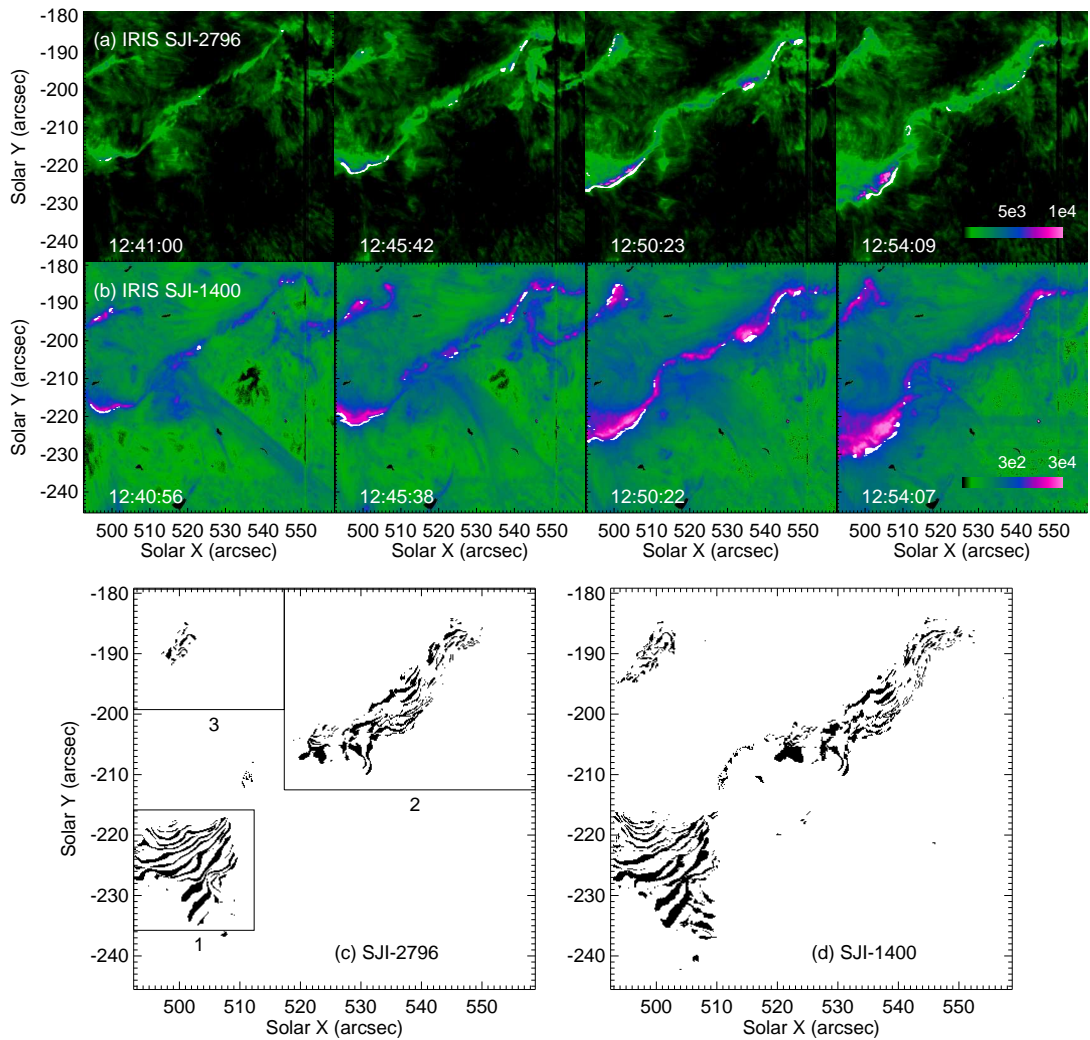}
\caption{Top: Snapshots of newly brightened ribbon fronts superimposed on ribbon images. (a) Ribbon fronts (white symbols) derived from the SJI-2796 images (brightness colored in linear scale, in units of counts) for threshold $N = 8$. (b) Ribbon fronts (white symbols) measured from the SJI-1400 images (brightness colored in normal logarithmic scale, in units of counts per second) for threshold $N = 100$. 
Bottom: Newly brightened ribbon fronts determined from (c) SJI-2796 and (d) SJI-1400. For clarity, the ribbon fronts are displayed at every third time increment (84~s cadence). The rectangular boxes (left) indicate Regions Of Interest (ROIs) 1, 2, and 3, in which the UV ribbon width and integrated UV counts are measured in the analysis. The field of view of all panels is the same as in Figure~\ref{fig:overview}c,d and the displayed heliographic coordinates represent the position of the flare region at 12:40:16~UT.
\label{fig:ribbon}}
\end{figure}

The flare-ribbon fronts are the collection of newly brightened pixels at each time. Figure~\ref{fig:ribbon} shows the time series of flare ribbons observed in (a) SJI-2796 and (b) SJI-1400, with the ribbon fronts denoted by the white markers. Panels (c) and (d) display the ribbon fronts at every third time frame for the SJI-2796 sequence with $N = 8$ and the SJI-1400 sequence with $N = 100$, respectively. Each strand in these two panels
outlines the positions of the newly brightened ribbon fronts during the 84-s interval between every third time frame. The {\em IRIS} SJIs fully cover the ribbon in the positive magnetic field. As the flare progresses, the ribbon spreads away from the PIL in the southwest direction.


It is generally accepted that the ribbon front delineates the intersection with the chromosphere of the magnetic field's separatrices, or quasi-separatrix layers, and that the spread of the ribbon front reflects the dynamical evolution of these structures due to reconnection in the corona \citep[e.g.][]{Sturrock1966, Demoulin1993, Savcheva2015}. The high-resolution {\em IRIS} SJIs reveal the fine-scale structure of the ribbon fronts. At any given instant, the newly brightened ribbon pixels form a curve that has finite width. During the flare evolution, the shape of this curve varies, and the overall width of the ribbon front varies. At any instant, the width also varies along the ribbon front. These variations of the ribbon width cannot be captured well in images with lower resolution, such as those obtained by AIA. Yet, the width of the ribbon front is thought to reflect the fine-scale structure and dynamics in the direction perpendicular to the coronal reconnecting current sheet \citep{Forbes1984,Hesse2005}. 

\begin{figure}[ht!]
\includegraphics[height=6cm]{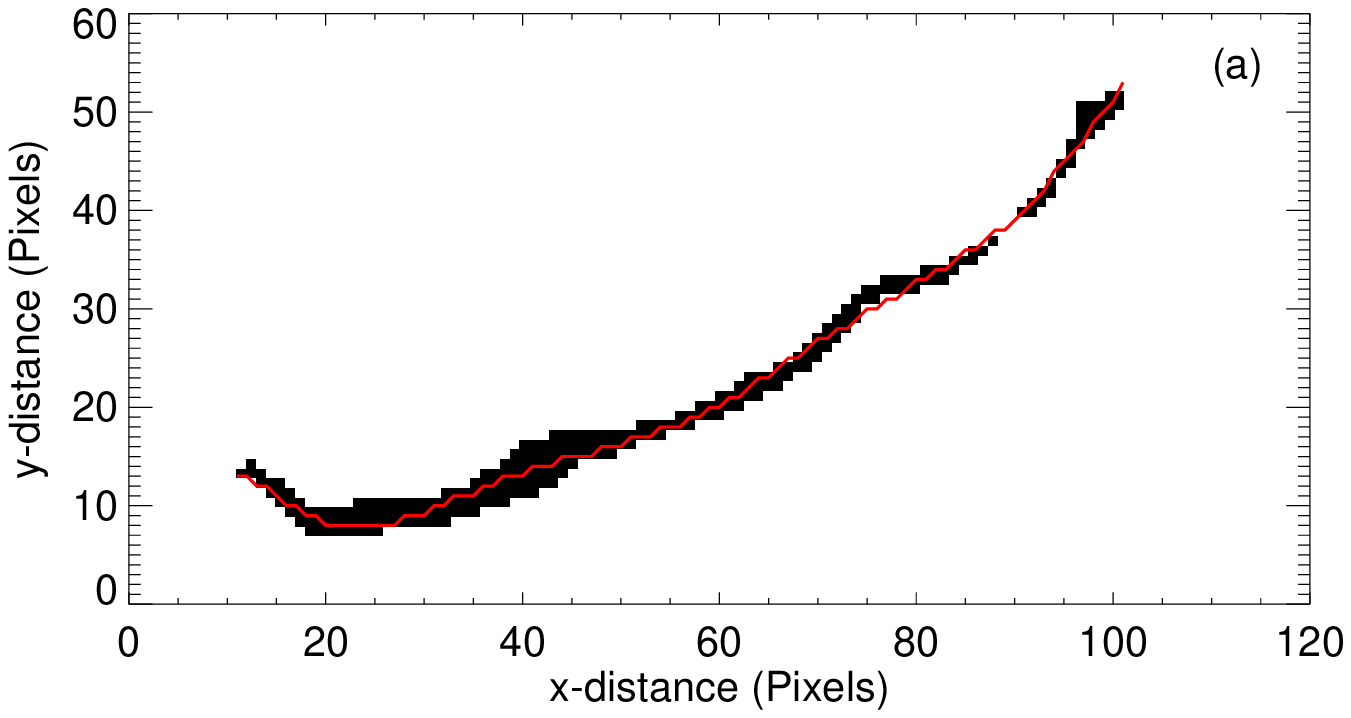}\includegraphics[height=6cm]{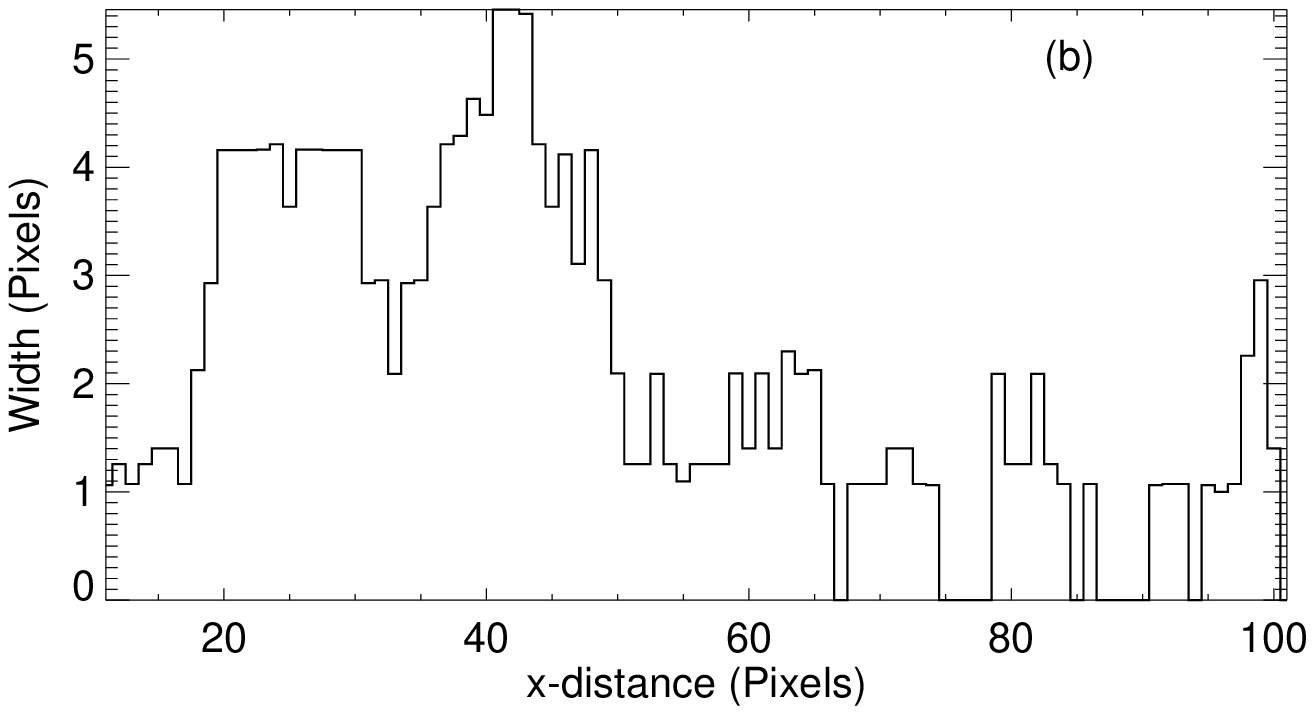}
\caption{(a) Image of a ribbon front in Region of Interest 1 (ROI-1; Figure \ref{fig:ribbon}c) as observed at $t$ = 12:48~UT in SJI-2796 for threshold $N = 8$ during the rise phase. The red line is the fitted curve along the ribbon front. (b) Width of the ribbon front as a function of position along the ribbon in the x-direction.}
\label{fig:measure1}
\end{figure}

To characterize the spatio-temporal variation of the spreading ribbons, we measured the width $\delta$ of the ribbon fronts at all time frames from $t$ = 12:34 to 12:57~UT.\footnote{After 12:57~UT, an ejection took place and part of the flare brightening no longer exhibited a coherent ribbon structure, so we did not measure the ribbon width thereafter.} We determined the points used to measure the perpendicular extent of the ribbon at each time frame by fitting the shape of the ribbon along its length to an 8th-degree polynomial. Figure~\ref{fig:measure1}a shows the polynomial curve fitted to a segment of the ribbon front as observed at $t$ = 12:48~UT in SJI-2796. 
The ribbon front width $\delta$ is calculated by finding the length of the shortest line that intersects the curve and stays within the ribbon-front structure at that time. Figure~\ref{fig:measure1}b shows the results after measuring $\delta$ at each point along the curve. At this time, the largest width is about 4-5 {\em IRIS} pixels, clustered toward the leftmost end of the ribbon front.

\begin{figure}[ht!]
\centering
\includegraphics[width=9cm]{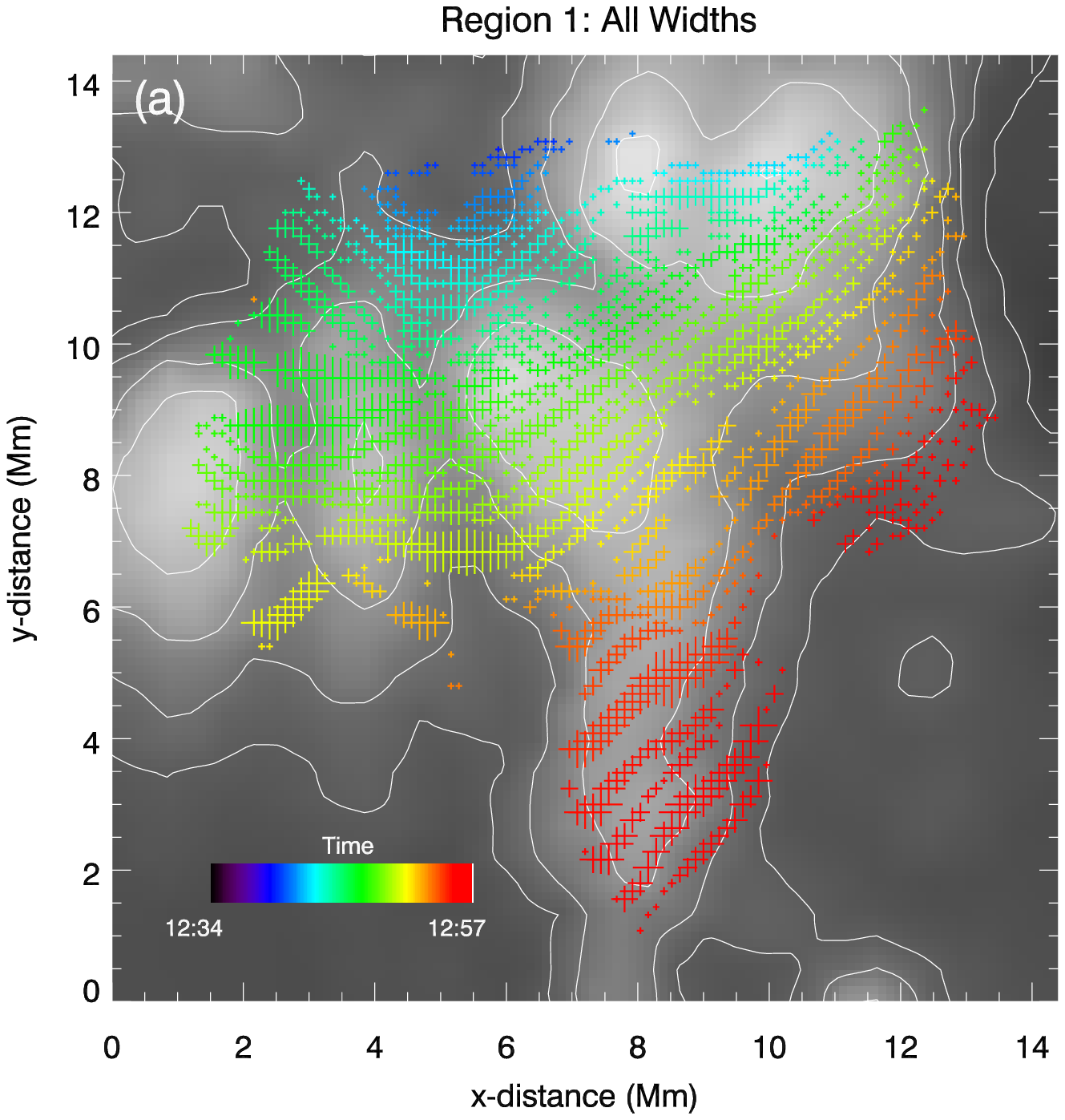}\includegraphics[width=9cm]{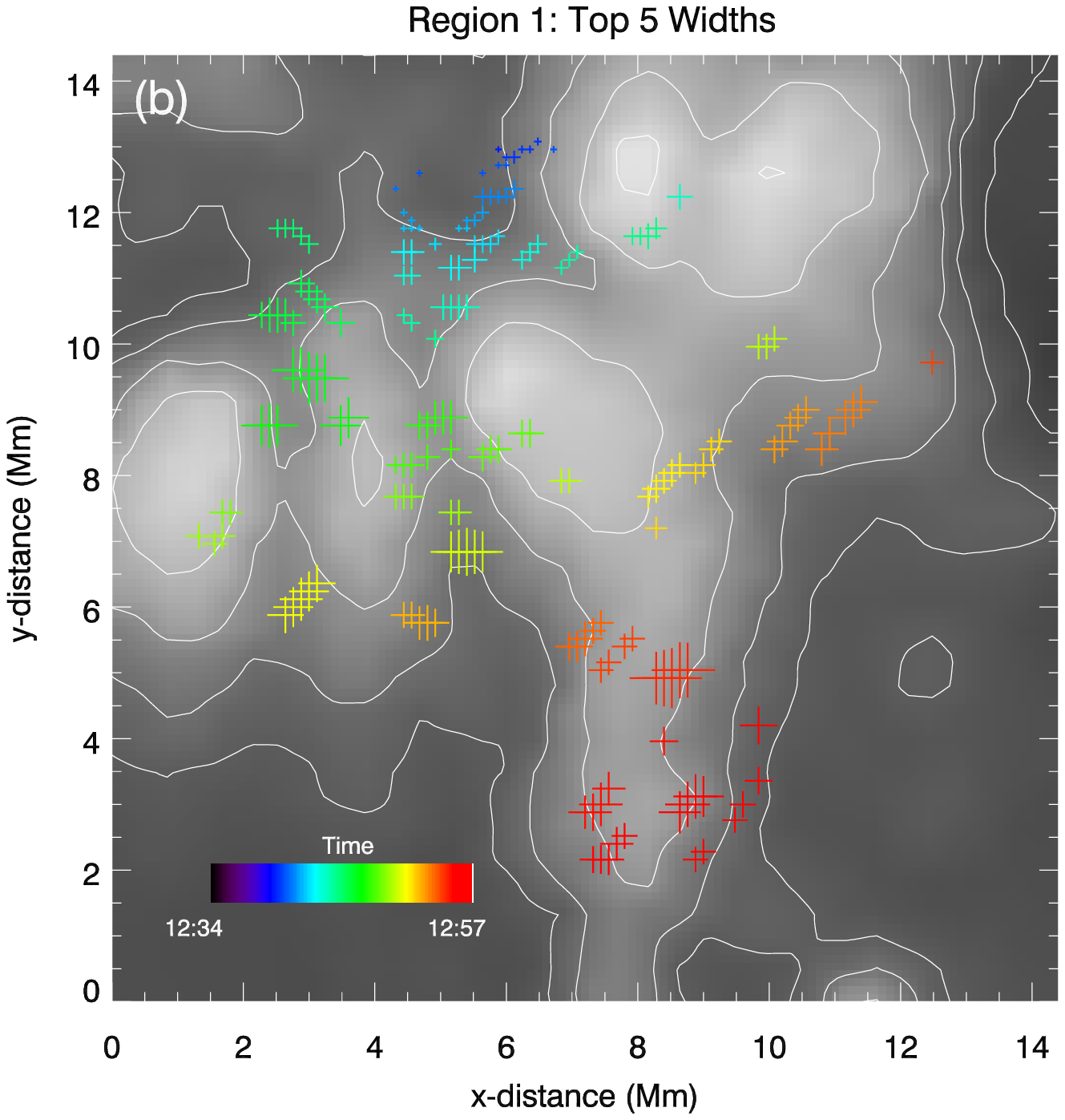}
\includegraphics[width=9cm]{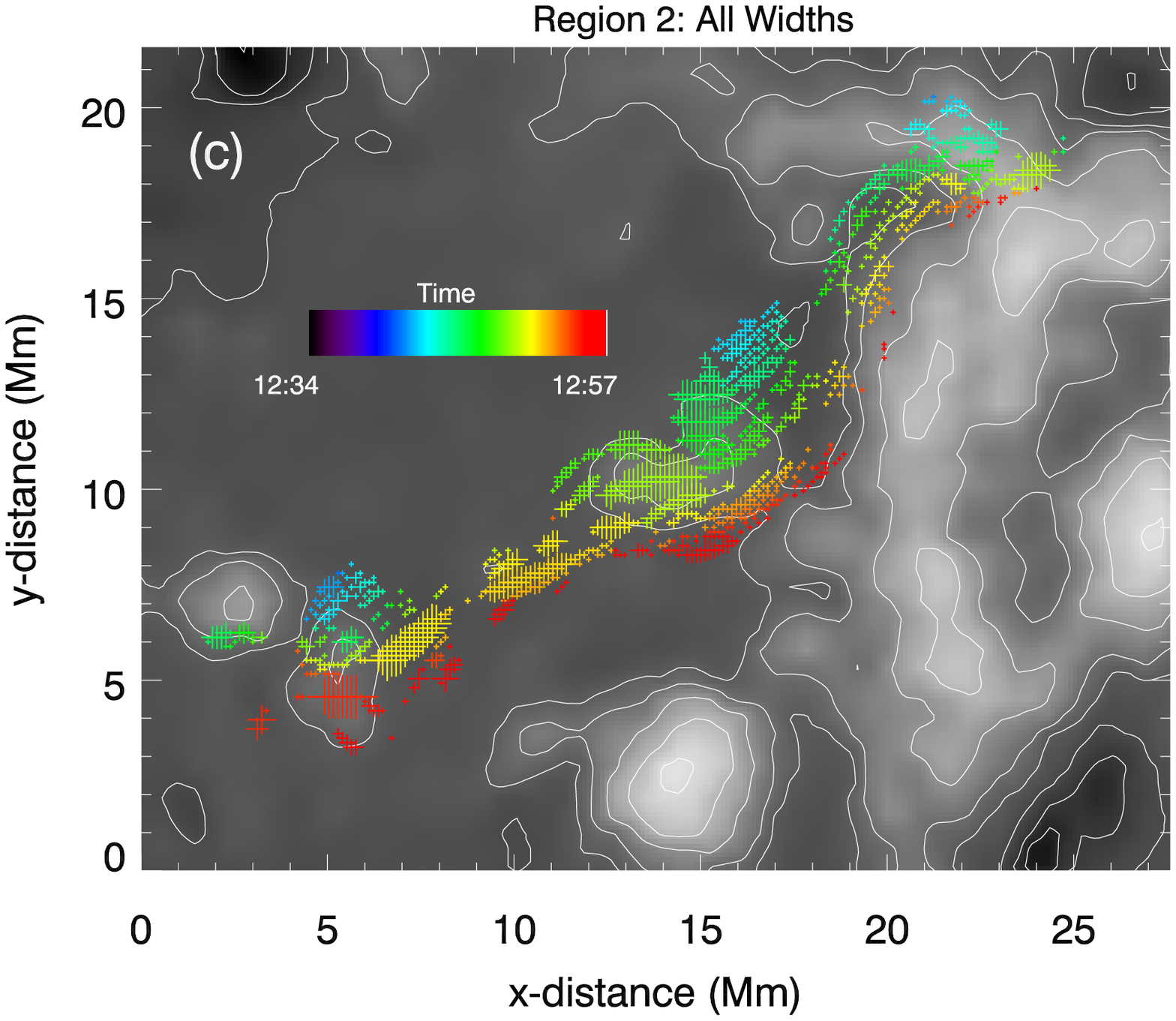}\includegraphics[width=9cm]{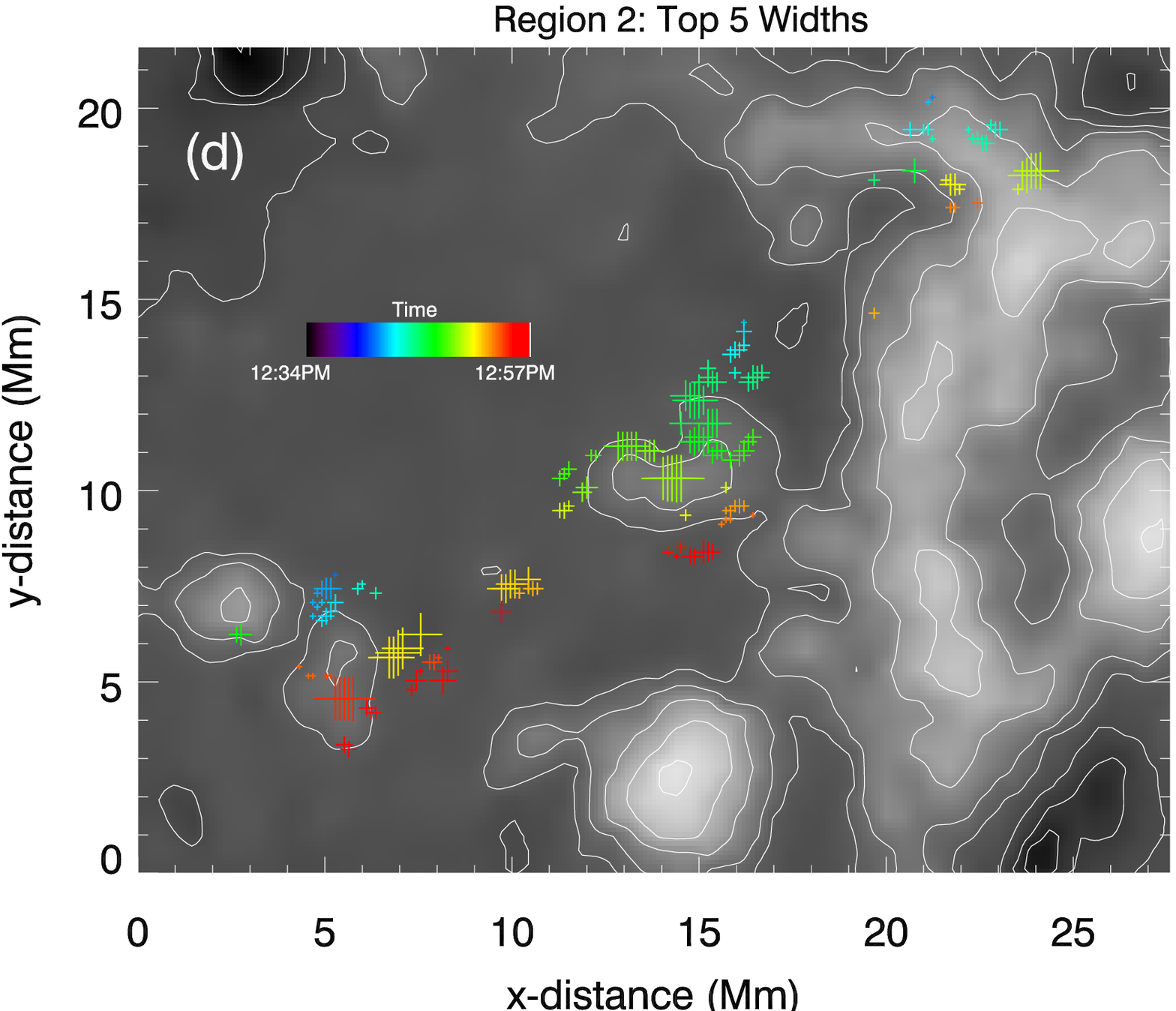}
\caption{Temporal evolution of the ribbon fronts measured with the SJI-2796 at $N = 8$. The symbol color indicates the time of brightening (see color bar); the symbol size is proportional to the width of the ribbon. The magnetogram $B_r$ of the region is in the background of each image with contour levels $\pm 0, 200, 400, 800, 1200, 1600$ Mx~cm$^{-2}$. ROI-1: Location at each time of (a) all widths and (b) top 5 widths. ROI-2: Location at each time of (c) all widths and (d) top 5 widths. Note the patchy structures in (b,d) compared to (a,c). 
The fields of view of the ROIs in these panels are indicated in Figures~\ref{fig:overview}b and \ref{fig:ribbon}c.
}
\label{fig:zebra}
\end{figure}

Within the field of view of the SJI, flare ribbons occupy three rather distinct regions, which we denote Regions of Interest (ROIs) 1, 2, and 3 and which are marked in Figures~\ref{fig:overview}b and~\ref{fig:ribbon}c. We evaluate the ribbon width in ROI-1 and ROI-2, the regions that exhibit the most coherent ribbon structure.\footnote{ROI-3 covers a part of the flare that is on the negative-polarity side of the PIL and is mostly excluded from the {\em IRIS} FOV.}
Figure~\ref{fig:zebra} shows the temporal and spatial distribution of all ribbon widths $\delta$ for ROI-1 (panel a) and ROI-2 (panel c) using SJI-2796 with $N =8$. 
In both regions, the general progression of the ribbon front is away from the PIL, from the northeast toward the southwest. Many portions of the fronts are coherent over distances of 1~Mm or greater; others are more fragmented. However, the ribbon width is quite variable along the more extended portions of the front.
To illustrate this, the top 5 widths at each time step are shown in Figures~\ref{fig:zebra}b and \ref{fig:zebra}d for ROI-1 and ROI-2, respectively. The clustering and fragmenting of the top-5 widths at each time step reveal that the locations with the largest widths appeared in the form of compact clusters at multiple locations along the ribbons. 
These features are consistent with a very spatially and temporally intermittent, or patchy, process of reconnection occurring in the overlying coronal flare current sheet. 

\begin{figure}[ht!]
\centering
\includegraphics[width=6.5cm]{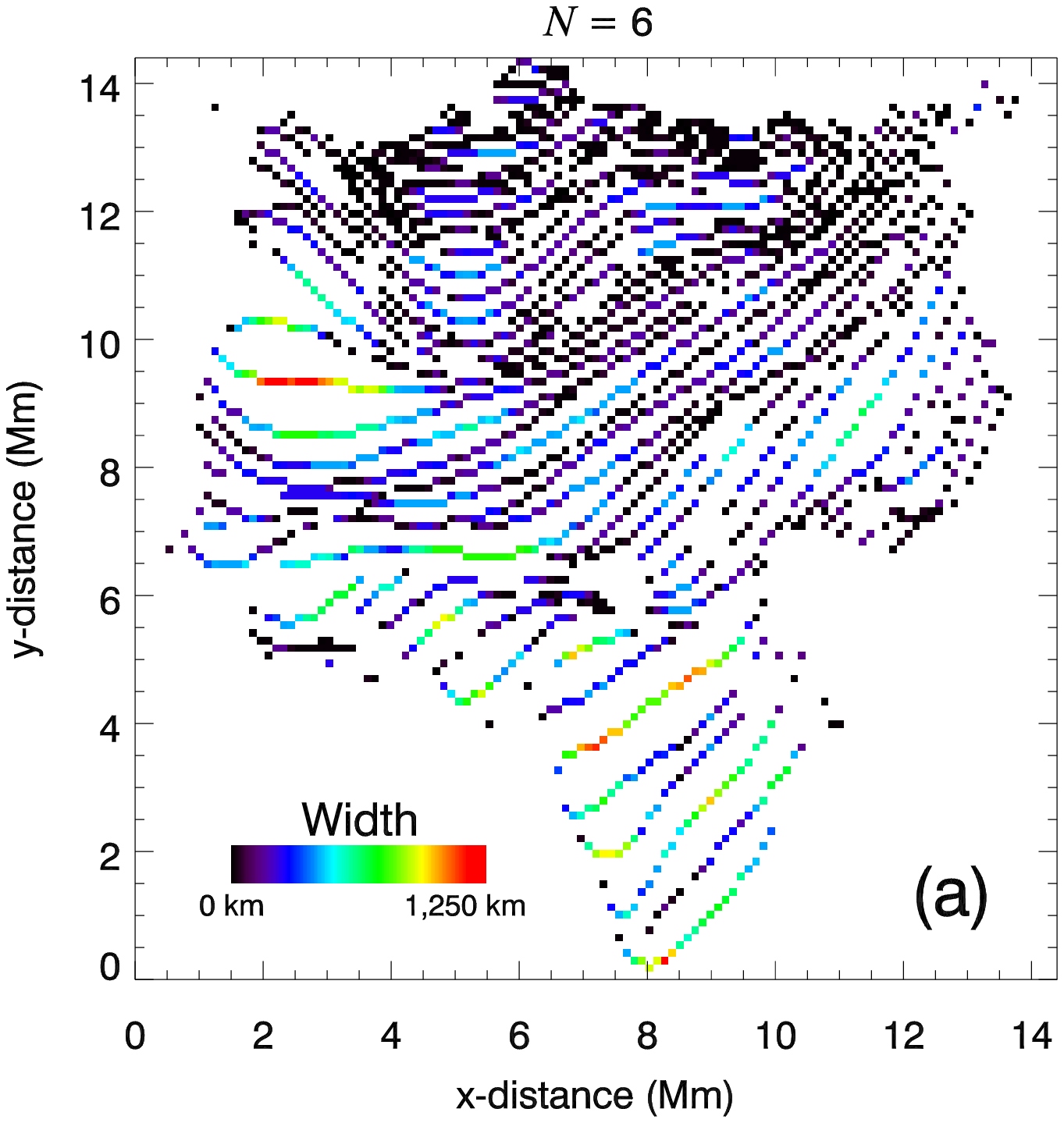}\includegraphics[width=10cm]{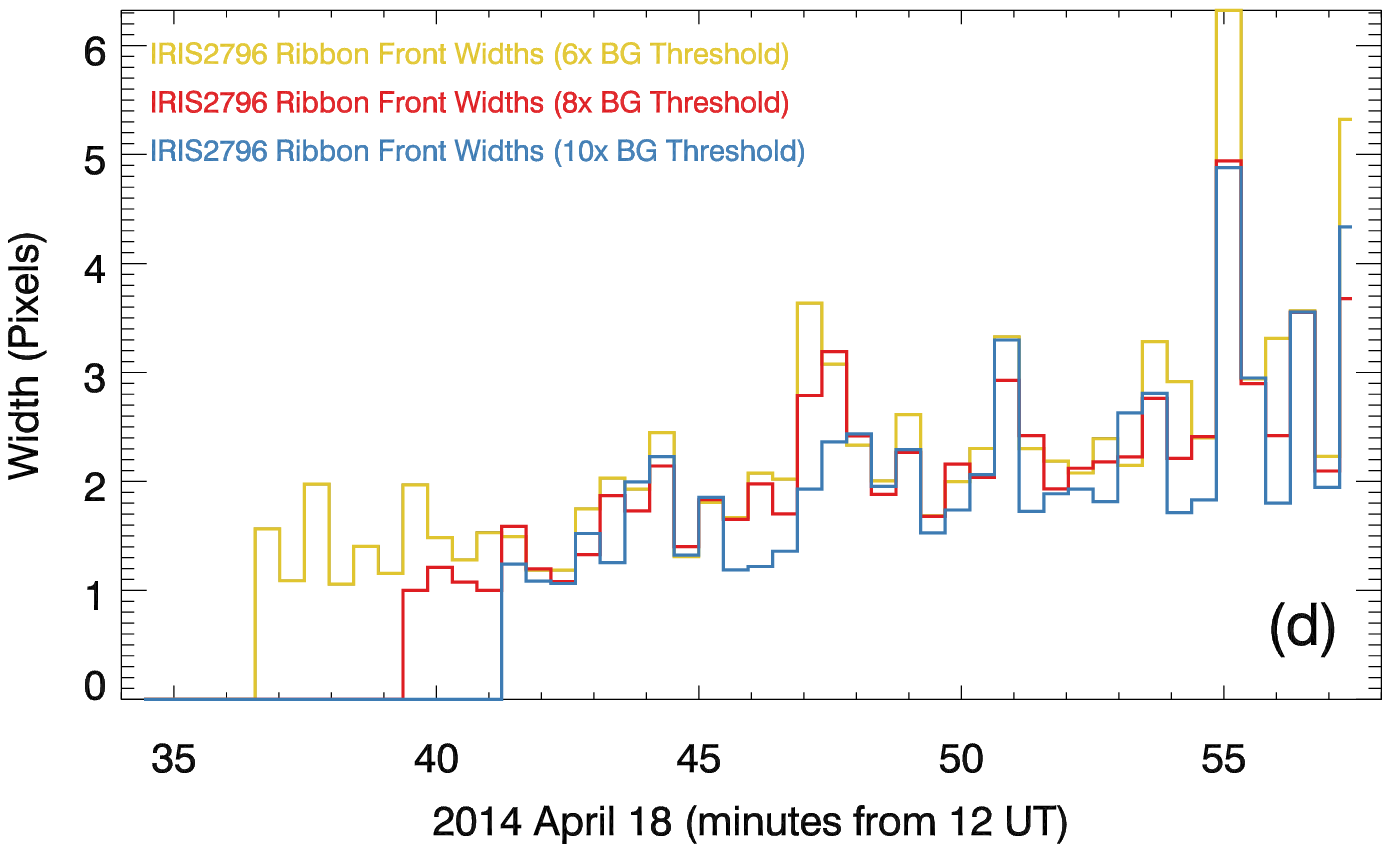}
\includegraphics[width=6.5cm]{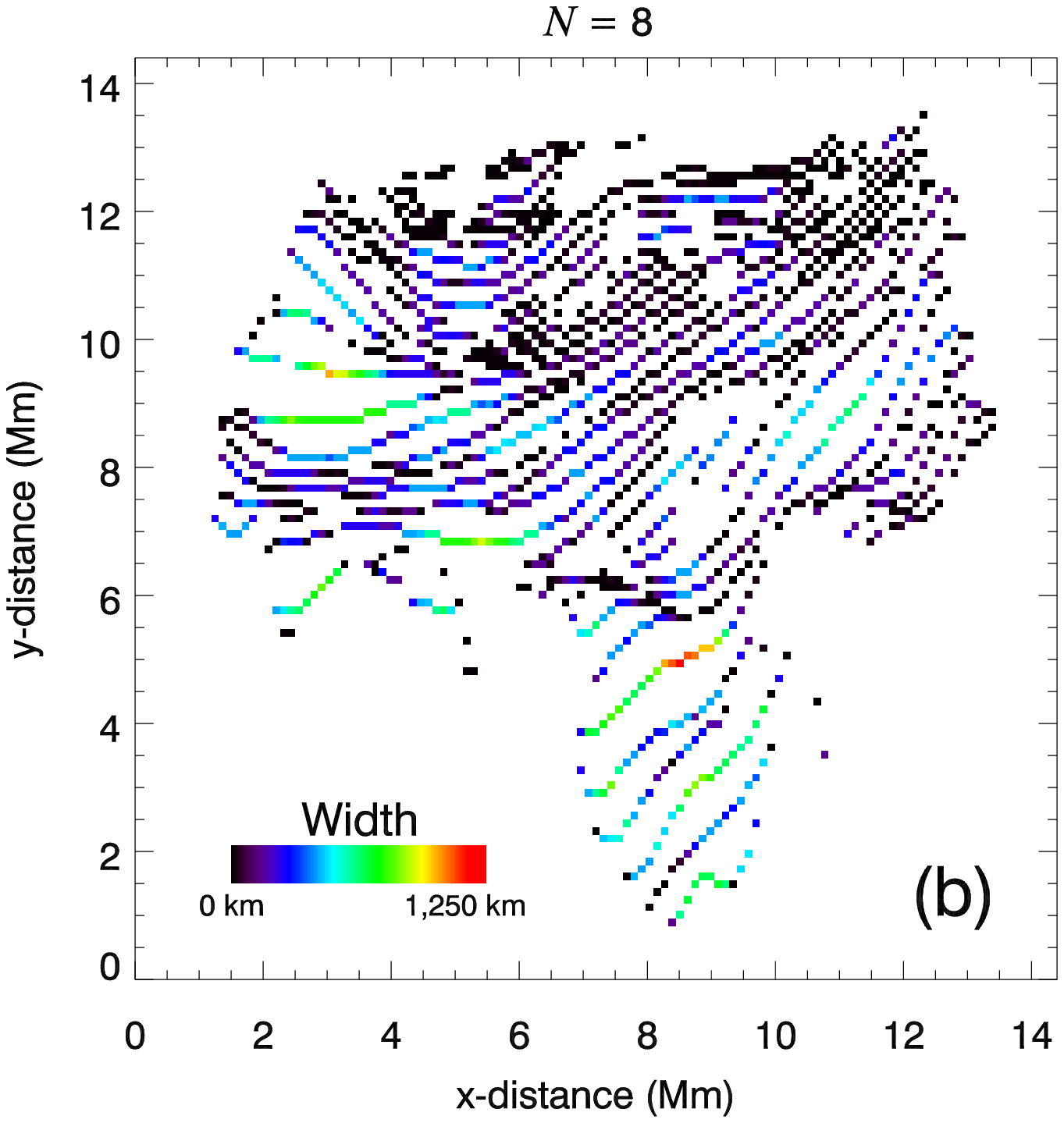}\includegraphics[width=10cm]{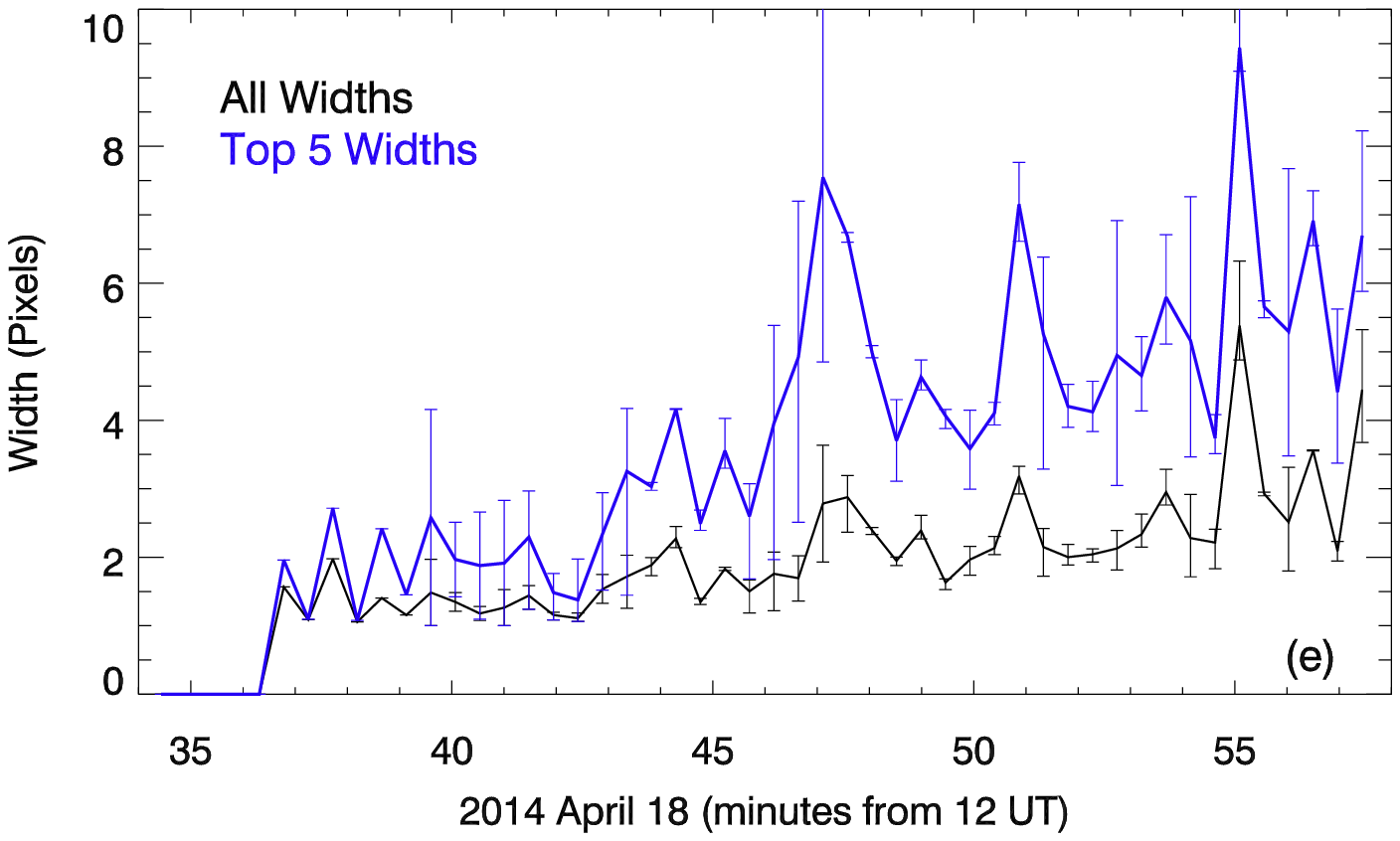}
\includegraphics[width=6.5cm]{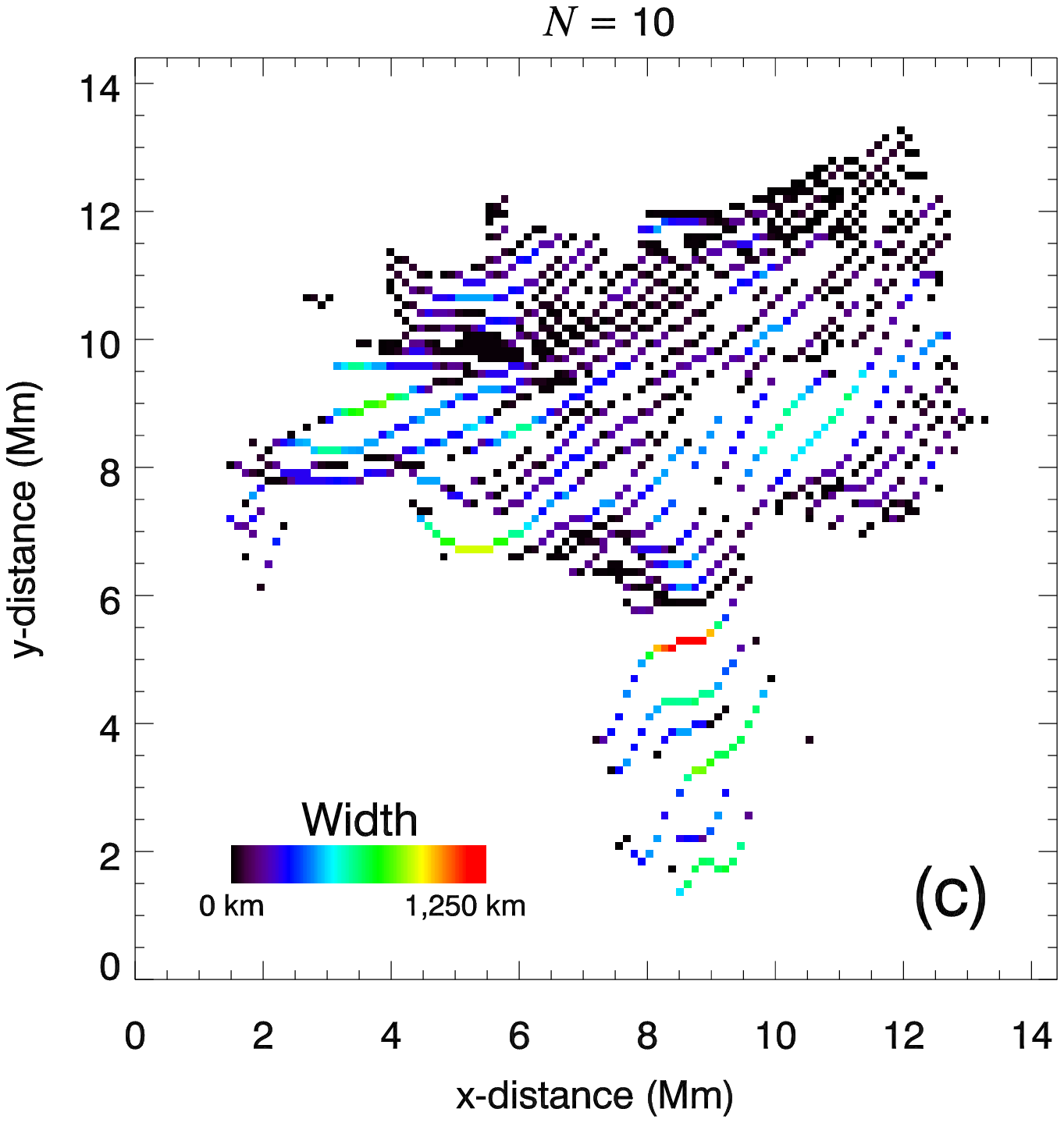}\includegraphics[width=10cm]{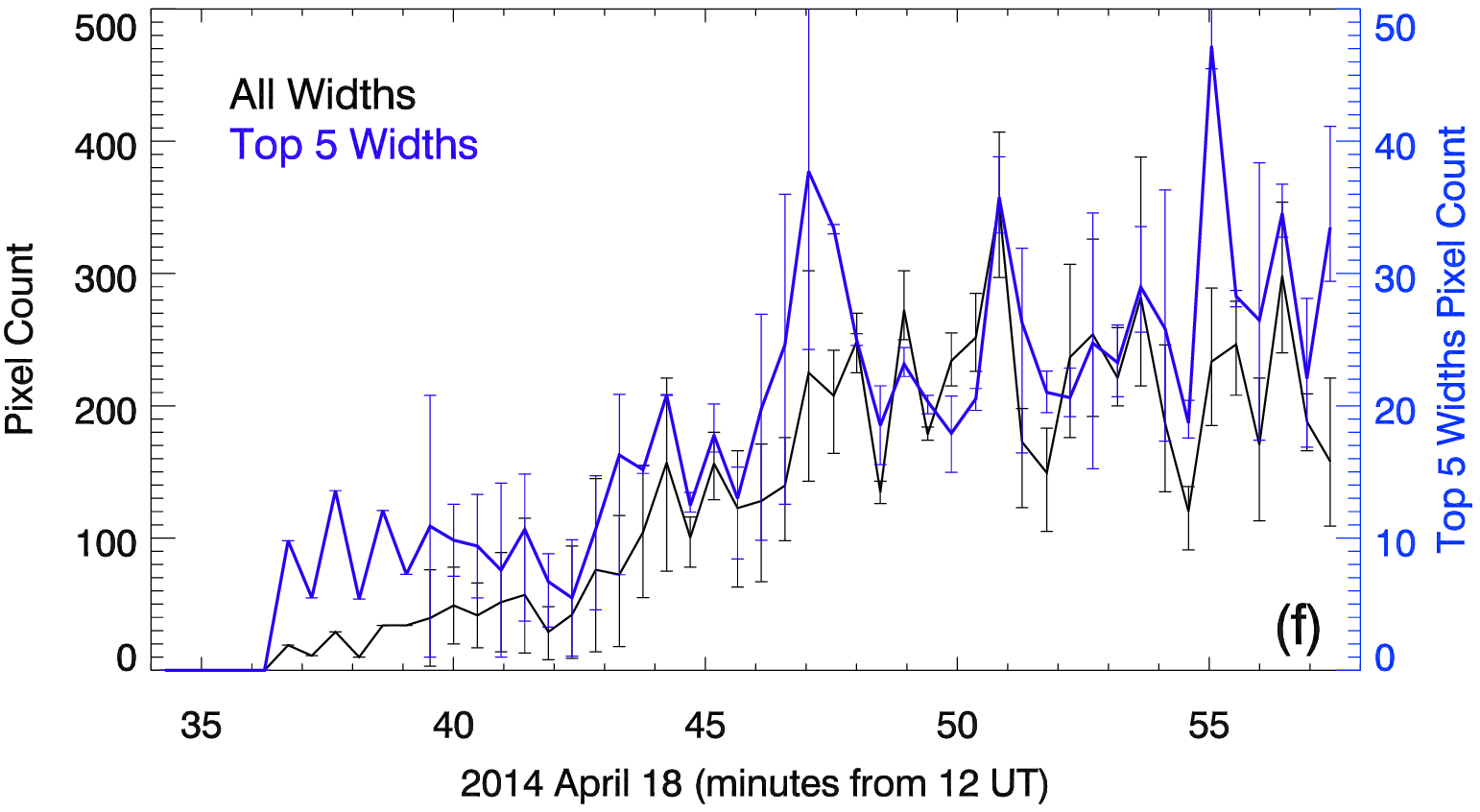}
\caption{Spatio-temporal pattern of the width of the ribbon fronts in ROI-1 derived from the SJI-2796 data with three thresholds $N$ = 6, 8, and 10 (a-c). Each panel shows the ribbon fronts at every time step; the color indicates the ribbon width. 
(d) Mean width $\langle \delta \rangle$ of the entire ribbon front for each threshold $N$ shown in (a-c). 
(e) Average and range of the measurements shown in (d) for all ribbon-front widths (black) and for only the top 5 ribbon-front widths at each time (blue). 
(f) Average and range of the total count of all ribbon-front {\em IRIS} pixels (black) and the ribbon-front {\em IRIS} pixels that make up the top 5 widths (blue). 
\label{fig:zebra2796}}
\end{figure}

\begin{figure}[ht!]
\centering
\includegraphics[width=6.5cm]{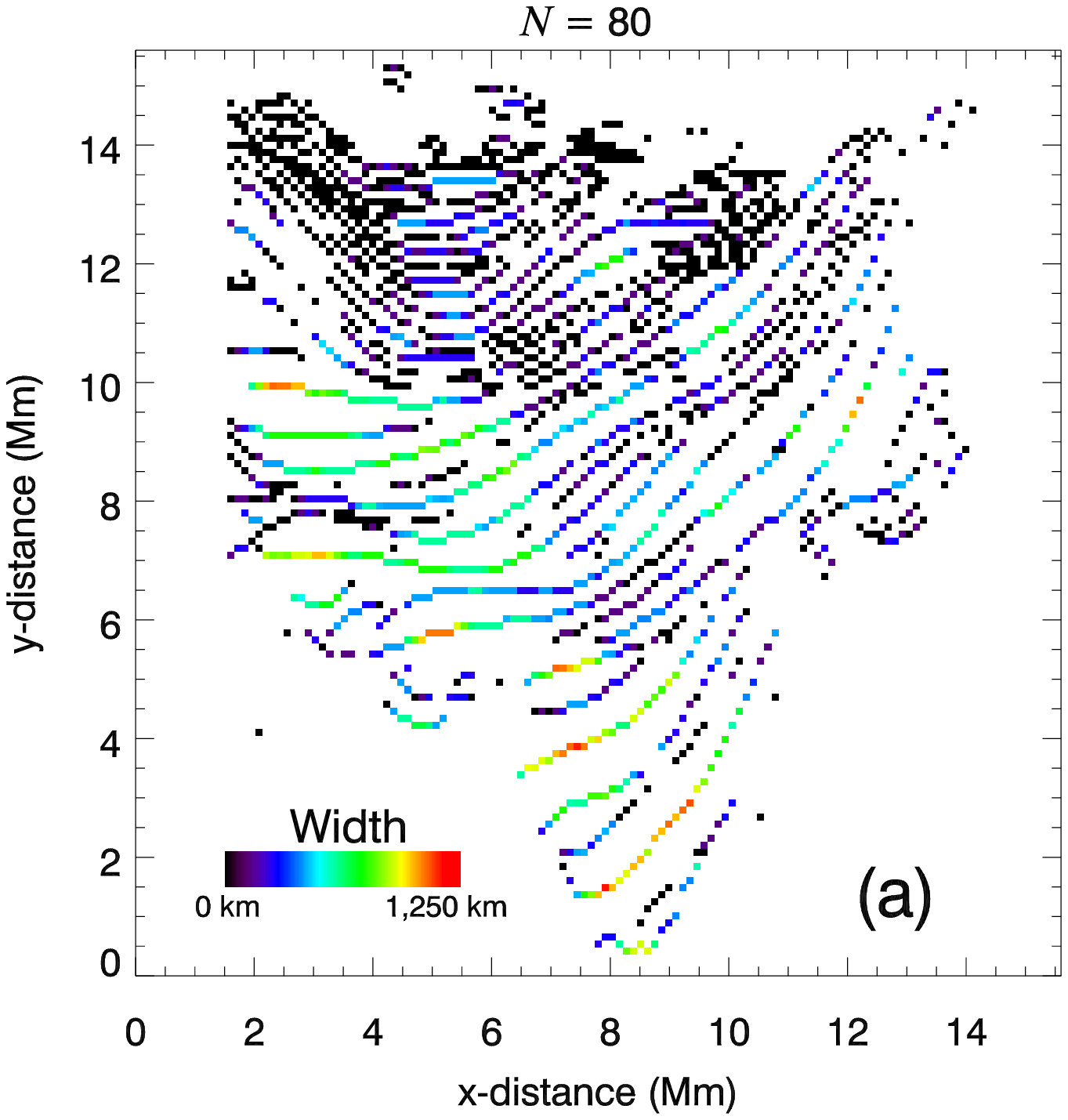}\includegraphics[width=10cm]{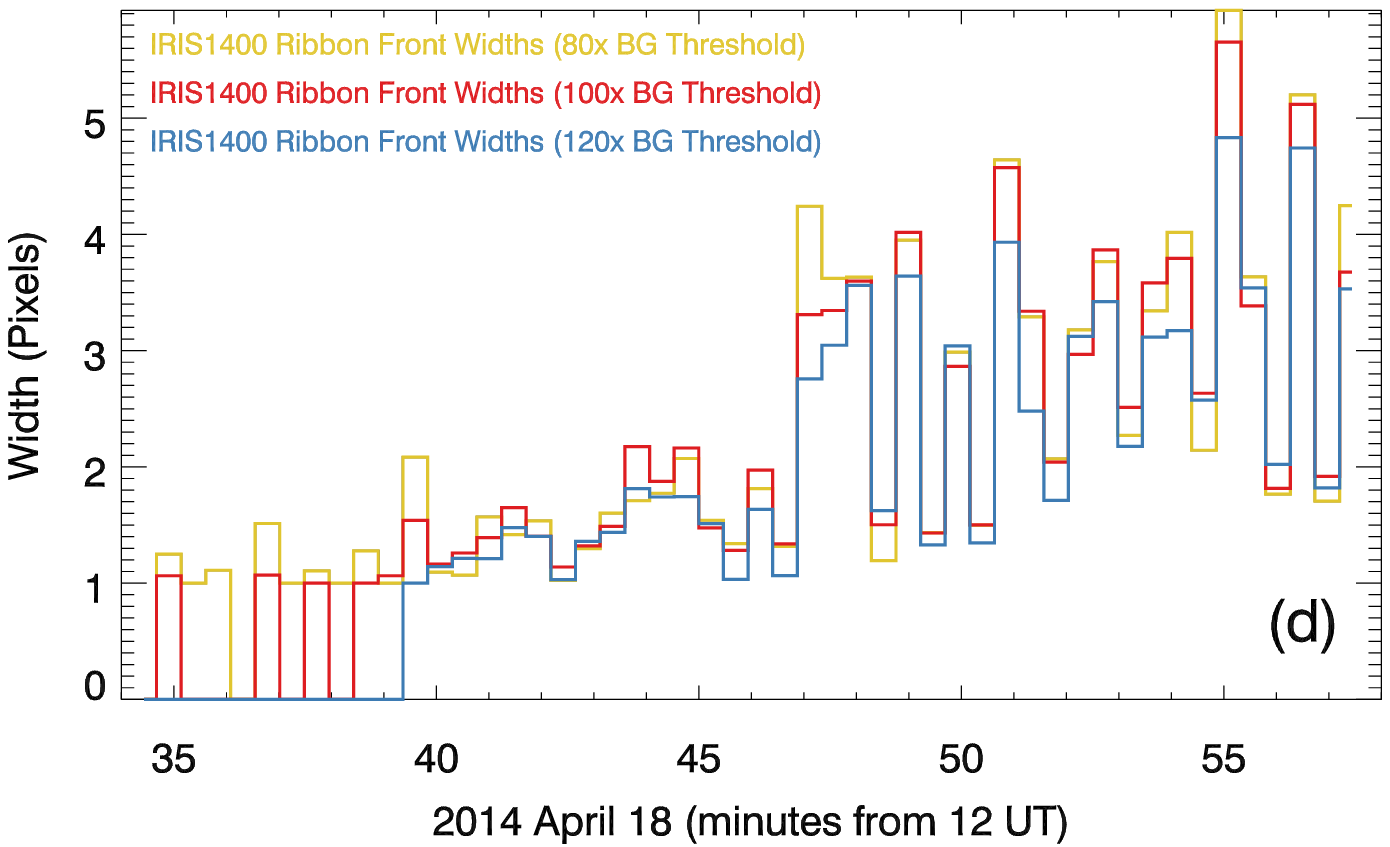}
\includegraphics[width=6.5cm]{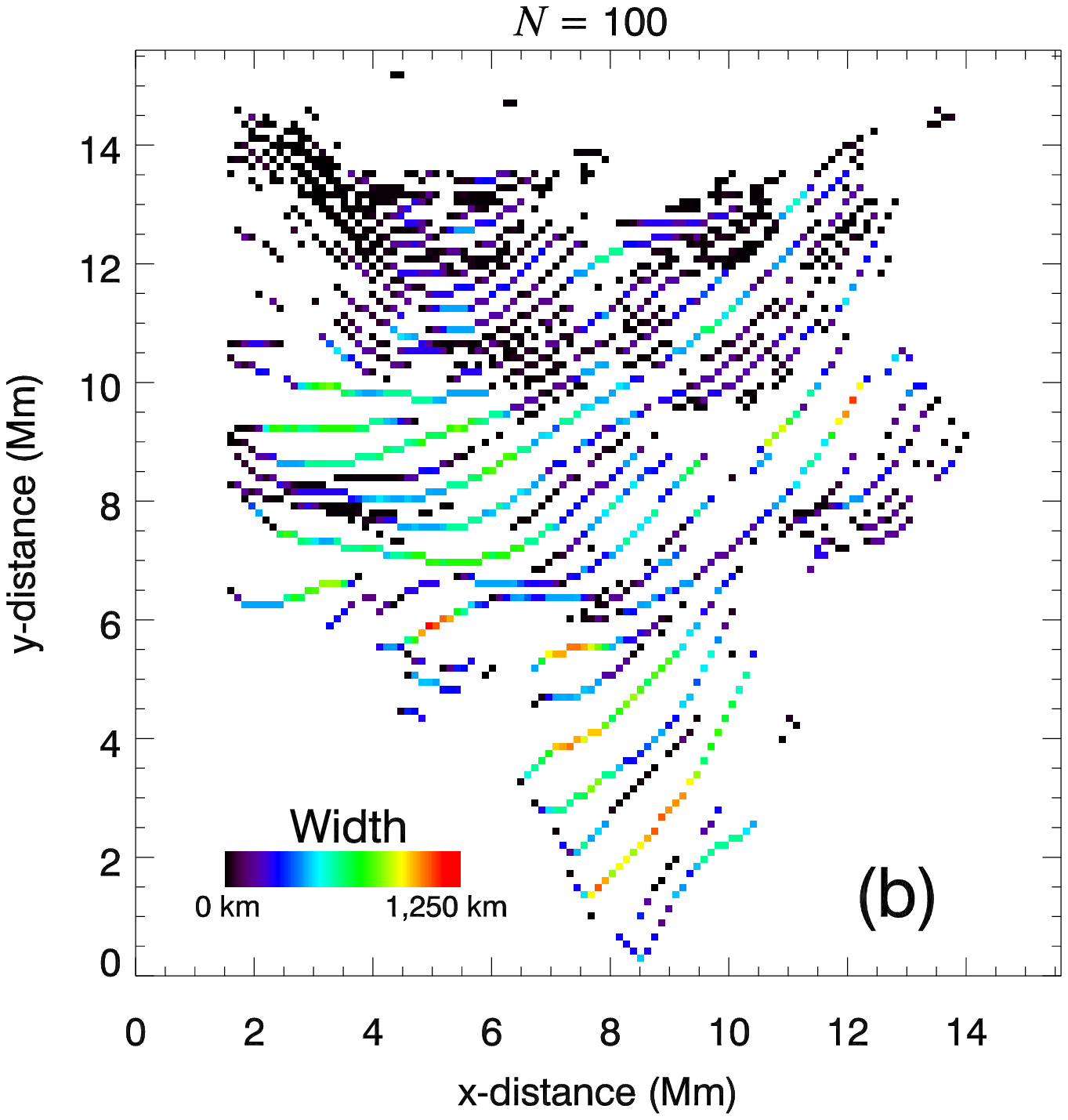}\includegraphics[width=10cm]{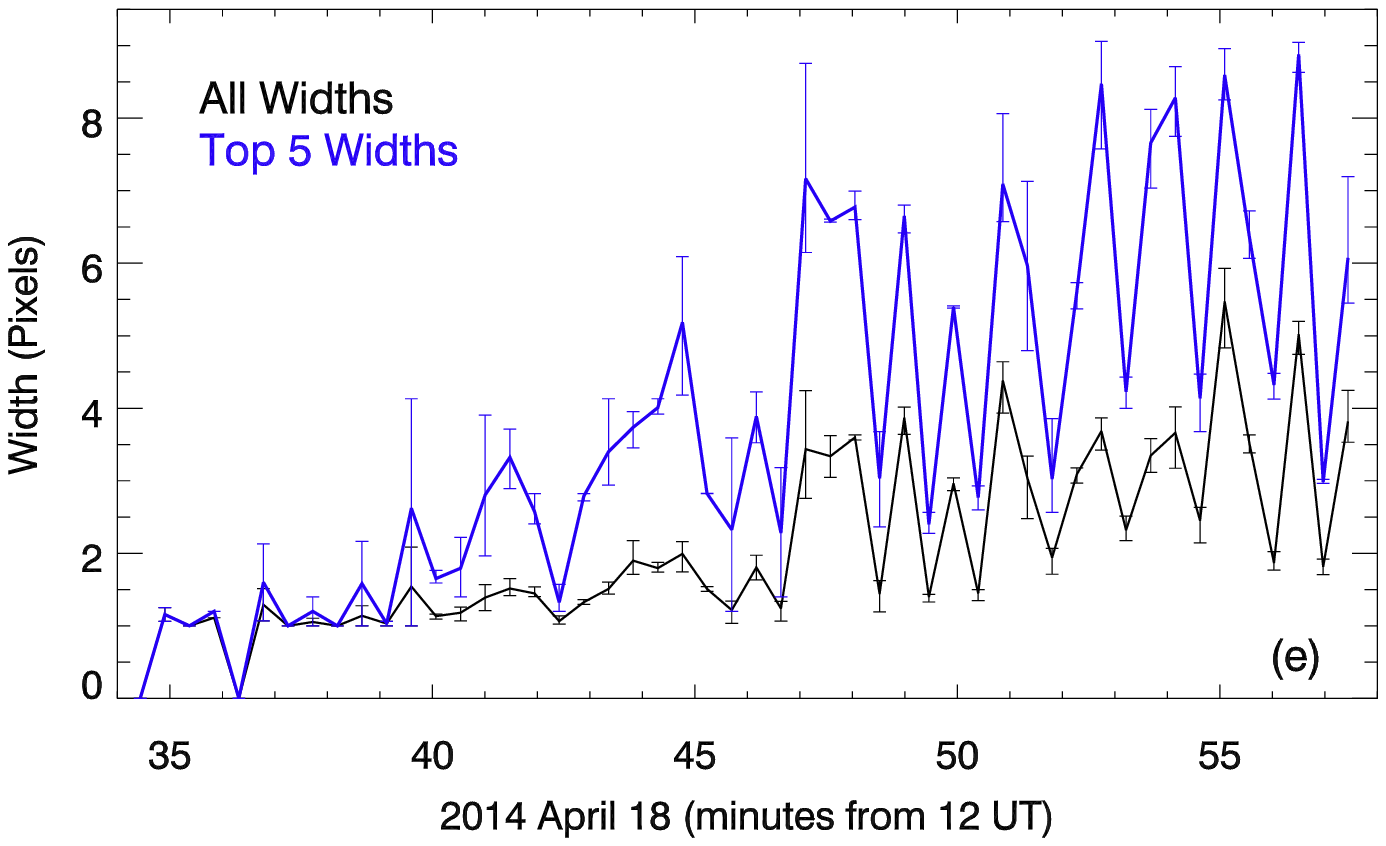}
\includegraphics[width=6.5cm]{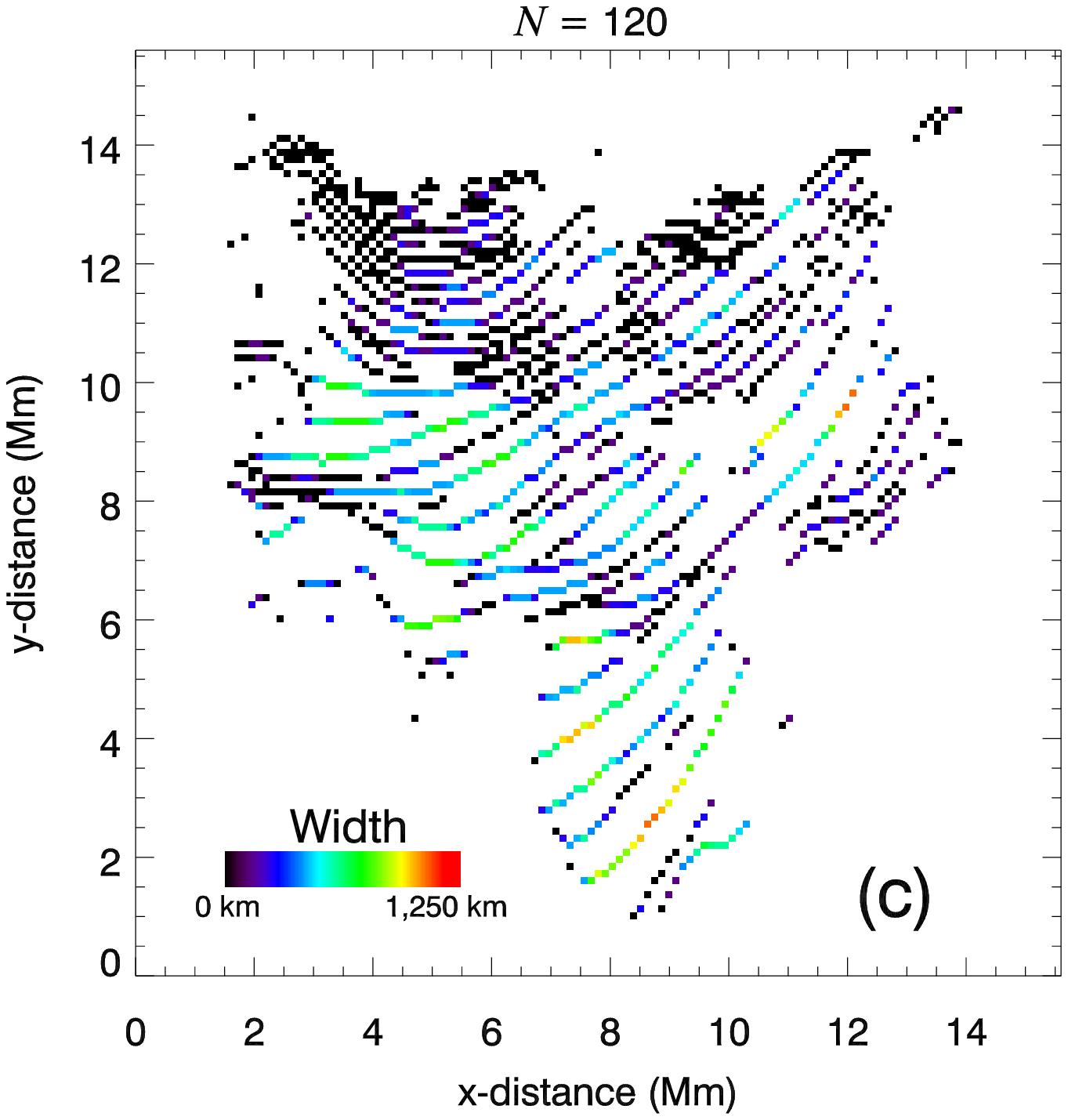}\includegraphics[width=10cm]{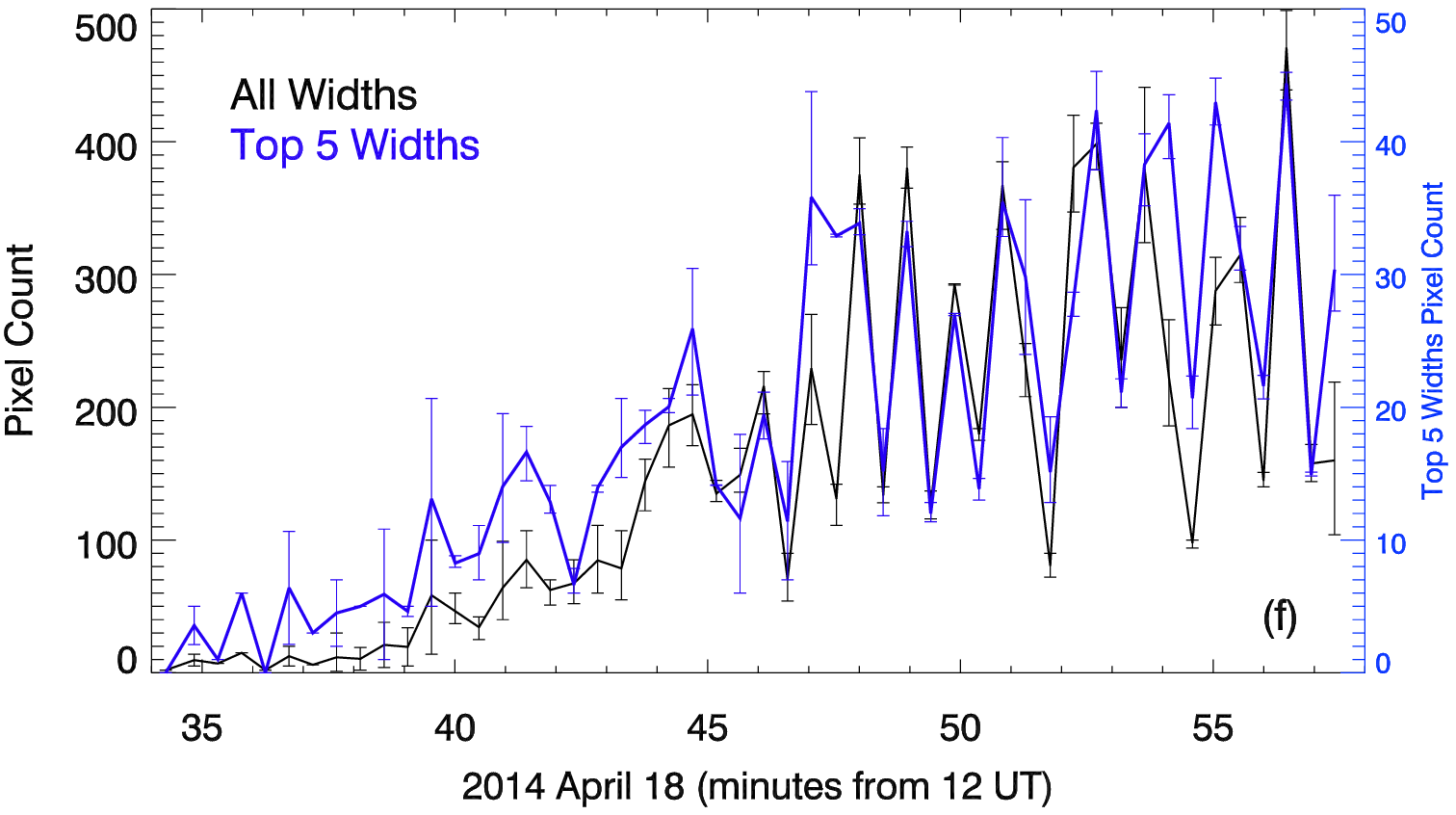}
\caption{Same as Figure~\ref{fig:zebra2796}, but using SJI-1400 data with three thresholds $N$ = 80, 100, and 120 (a-c). 
\label{fig:zebra1400}}
\end{figure}

The width measurements using different data sets and different thresholds give rise to quite similar patterns. We establish this in Figures~\ref{fig:zebra2796} and \ref{fig:zebra1400} by showing our measurements made in ROI-1 using the SJI-2796 data with $N$ = 6, 8, and 10 (Fig.\ \ref{fig:zebra2796}) and the SJI-1400 data with $N$ = 80, 100, and 120 (Fig.\ \ref{fig:zebra1400}). The overall trends are quite similar in the two data sets\footnote{The ribbon front widths presented in this paper are measured using the level-2 {\em IRIS} SJIs without correcting the stray light. We have also analyzed SJI-2796 images with the correction of the Point Spread Function (PSF) recently determined by the {\em IRIS} team {\citep[also see][]{Alissandrakis2018, Courrier2018}}, and the resultant spatio-temporal pattern of the ribbon fronts does not exhibit a substantial difference from those shown in Fig.~\ref{fig:zebra2796}.}; nevertheless, there are a few qualitative differences that are worth noting. 

Panels (a-c) in each case show three maps, differing only in the thresholds used to obtain the ribbon-front widths. As might be anticipated, the morphologies of the three maps are very similar within each data set and across the two data sets, as well. As the threshold is raised progressively, the distribution of ribbon-front pixels becomes more sparse and the boundaries of the region occupied by these pixels generally recede inward towards the center. These trends reflect the declining number of affected pixels due to the exclusion of those that have brightened insufficiently to reach the higher thresholds. Differences between the various maps clearly are quantitative, not qualitative, which indicates that our ribbon-front measurements are robust across different choices for the threshold values and for the wavelength/temperature of the emission. This close correspondence between the SJI-2796 maps (Fig.\ \ref{fig:zebra2796}a-c) and the SJI-1400 maps (Fig.\ \ref{fig:zebra1400}a-c) is not surprising, but it also is not trivial, given the substantial contrast in formation temperature of the two bandpasses, $T_{2796} \approx$ 10,000~K (chromosphere) and $T_{1400} \approx$ 63,000~K (transition region), and the vastly different dynamic ranges of the enhanced emission in these two passbands in response to energy deposition in the lower atmosphere (see the comparison of the single-pixel light curves in Figure~\ref{fig:lgtcv}). Indeed, this contrast seems to be responsible for the main qualitative differences between the two sets of maps: the SJI-2796 ribbon fronts become both more sparse and less extended spatially at increasing thresholds at a much faster rate than do the SJI-1400 ribbon fronts. This is evident from the much greater contrast between panels (a) and (c) in Figure~\ref{fig:zebra2796} compared to Figure~\ref{fig:zebra1400}.  

Time histories of the ribbon widths measured in ROI-1 in these two passbands are displayed in Figures~\ref{fig:zebra2796}d and \ref{fig:zebra1400}d. The mean of the measured widths at thresholds $N$ = 6/80 (gold), 8/100 (red), and 10/120 (blue) from the SJI-2796/SJI-1400 data are shown. For the SJI-2796 case (Fig.\ \ref{fig:zebra2796}d), the thresholds are attained at progressively later times -- 12:37, 12:40, and 12:42~UT, respectively -- indicating that the ribbon fronts gradually build in strength, from lower intensity at earlier times to higher intensity at later times. Overall, the maximum widths at each threshold increase over time. but they also exhibit substantial fluctuations from one snapshot to another. Moreover, the maximum width is dominated by low-intensity fronts at certain times (e.g.,\ 12:47~UT) but by high-intensity fronts at other times (e.g.,\ 12:55~UT). These curves reveal considerable complexity and variability in the instantaneous structure of the ribbon fronts as the flare proceeds. The widths measured from the SJI-1400 data (Fig.\ \ref{fig:zebra1400}d) exhibit the same general features, but the variations among the different thresholds are substantially smaller. This is consistent with the lesser contrast among the ribbon front maps shown in panels (a-c) for the latter data set. 

From these measurements, we derive mean values and ranges of variation of the ribbon-front widths, $\delta$, for the two bandpasses. 
 Panel (e) shows the evolution of the ribbon-front widths derived using all measurements ($\langle \delta\rangle$; black) and using only the measurements 
of the top 5 width at each time ($\langle \delta_{5} \rangle$; blue).
The two pairs of curves follow similar trends, generally increasing but with substantial fluctuations from the beginning to the end of the flare. The SJI-1400 widths (Fig.\ \ref{fig:zebra1400}e) fluctuate rather more rapidly and vigorously than the SJI-2796 widths (Fig.\ \ref{fig:zebra2796}e) during the late phase, after about 12:45~UT. During the early phase, before about 12:45~UT, the wider ribbon-front segments ($\langle \delta_{5} \rangle$) fluctuate strongly and exhibit intermittent dropouts in both passbands. The mean width of all segments reaches about 3 {\em IRIS} pixels late in the flare, whereas the mean of the wider segments reaches about 8 {\em IRIS} pixels. {It is also noted that, although the exposure of SJI-1400 images varies during the flare, the variation of the ribbon front width measured in this image set tracks very well the variation measured in SJI-2796 images obtained with a constant exposure. Therefore, these ribbon width variations are primarily produced by the progress of the flare, rather than by other effects such as the varying exposure.}

Figures~\ref{fig:zebra2796}f and \ref{fig:zebra1400}f show the total number of ribbon-front pixels and the number of ribbon-front pixels that contribute to the top 5 width measurements at each time. 
Both curves 
reach broad peaks at about 12:48~UT and stay at that level, albeit with fluctuations similar to those in the ribbon-width curves. The total count of
pixels contributing to the top 5 widths amounts to 20\% of the total number of newly brightened ribbon front pixels in the early phase, and then drops to about 10\% after 12:48~UT.
This behavior indicates that the early ribbon brightening occurs predominantly in thin fronts that are distributed widely across the region, whereas the late brightening occurs increasingly in wider fronts at a more nearly uniform rate. {As described earlier in the text, the total number of the ribbon front pixels varies with the threshold. With the set of thresholds chosen in this study, when a higher threshold is used, the total count of the ribbon front pixels is reduced by 25\% for SJI-2796, and by 12\% for SJI-1400. The variation in the mean ribbon width caused by changing thresholds is insignificant.}

\subsection{Connection with UV Ribbon Brightness}
\label{subsec:uv}

\begin{figure}[ht!]
\centering
\includegraphics[width=9cm]{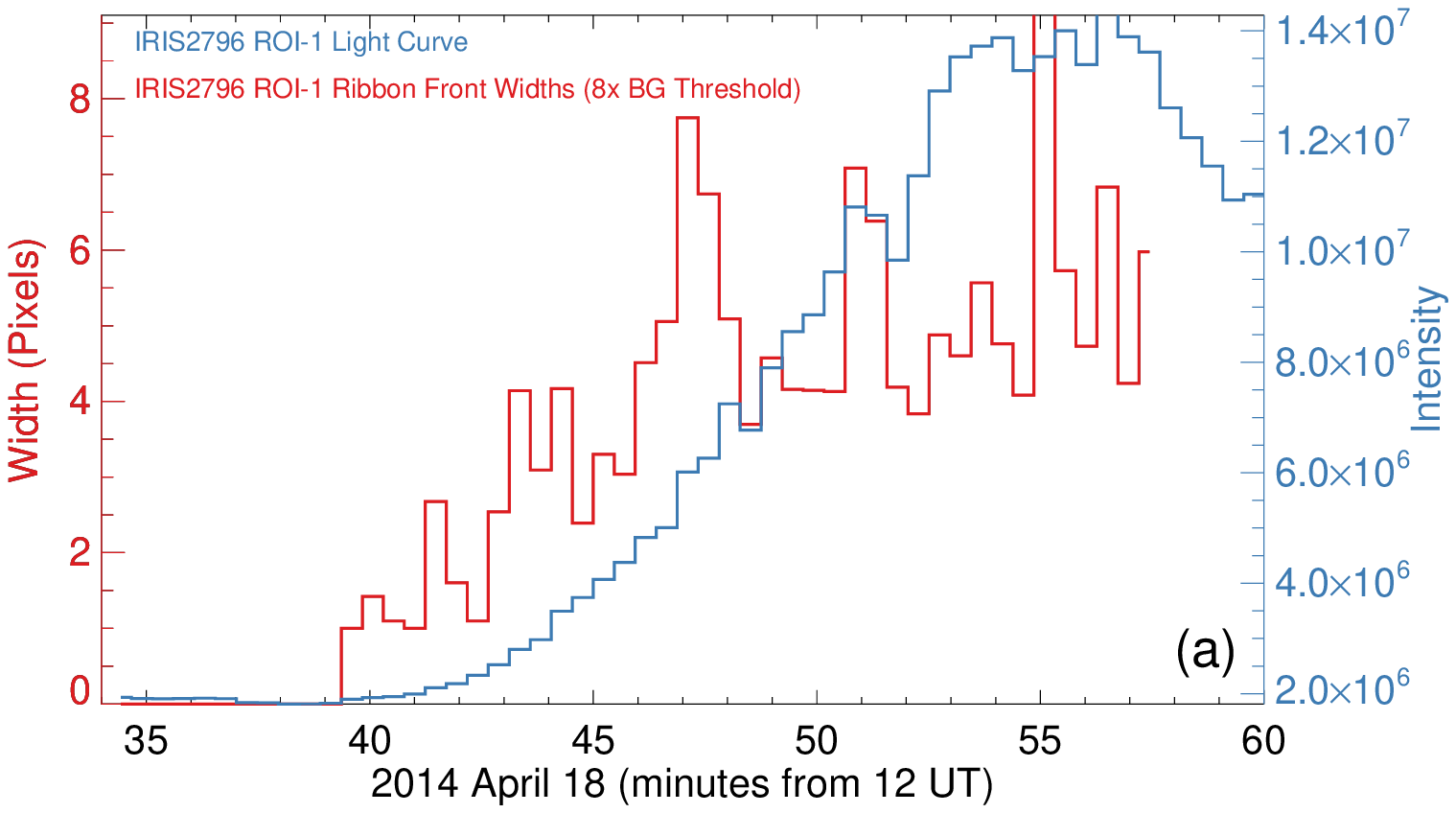}\includegraphics[width=9cm]{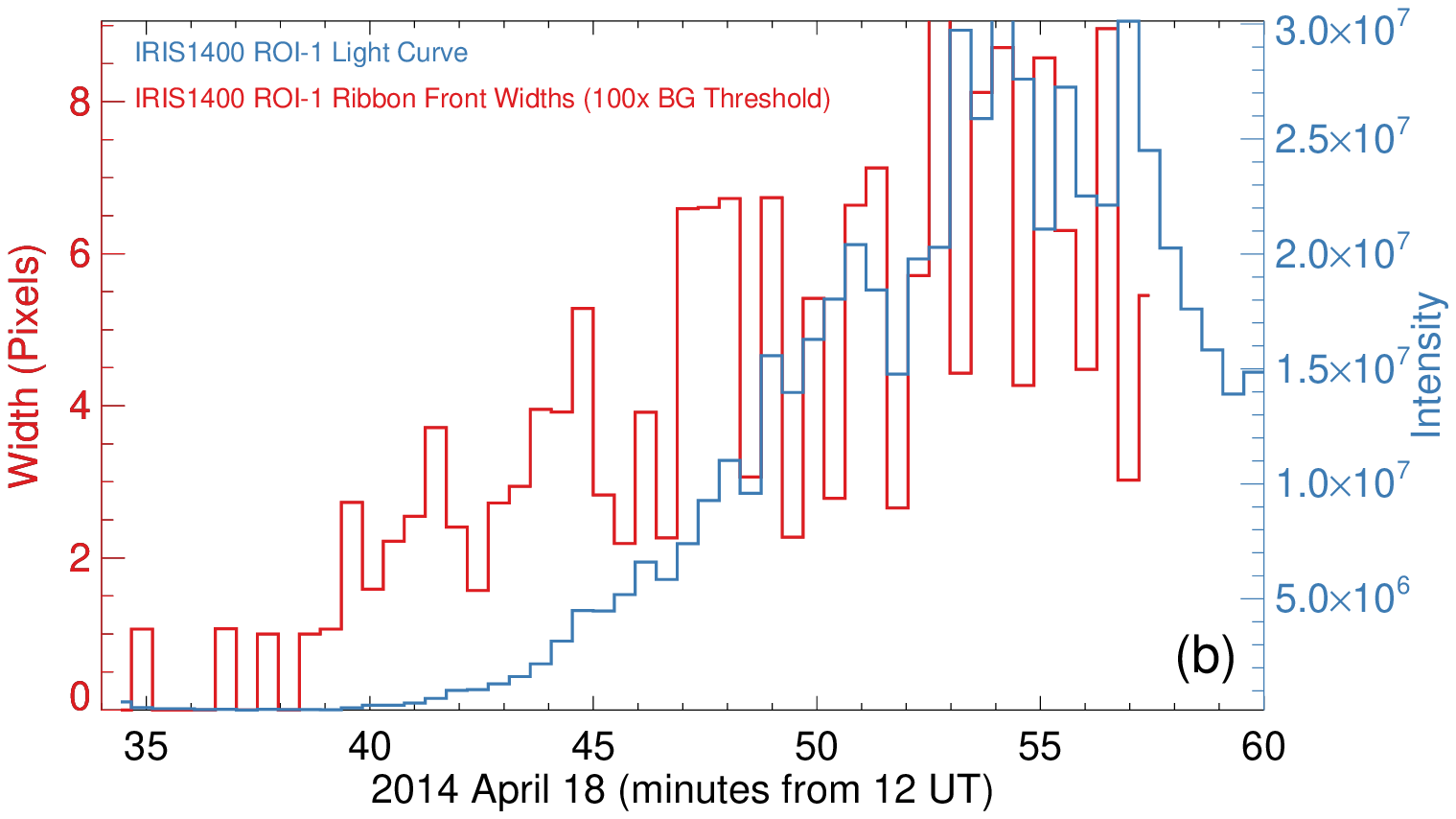}
\caption{Average of the top five ribbon front widths at each time (red) and integrated light curves (blue) in ROI-1 for (a) SJI-2796 (threshold $N$ = 8) and (b) SJI-1400 (threshold $N$ = 100). 
\label{fig:widths_lightcurves}}
\end{figure}

To explore further the relationship between the intensity of the ribbon emissions and the width of the ribbon segments, we compare these two quantities for SJI-2796 and SJI-1400 in region ROI-1 in Figure~\ref{fig:widths_lightcurves}. The total integrated light curves (above the background threshold) are shown in blue, and the ribbon-front widths averaged over the top five values at each time are shown in red. The results confirm the evidence presented in the preceding section: the ribbon brightenings begin early (about 12:40~UT) at very low intensity, when the light curve is negligibly small compared to its value at the flare peak time (about 12:55~UT). The widths grow rapidly to reach nearly their peak values by about 12:47~UT, when the intensity has yet to reach 50\% of its maximum. Thereafter, the width essentially plateaus, with fluctuations, while the intensity continues to climb until it tops out at about 12:55~UT and declines thereafter. The intensity peaks are accompanied by new, but only slightly higher, peaks in the ribbon-front widths. These qualitative behaviors of the ribbon widths and intensities are very similar for SJI-2796 (Fig.\ \ref{fig:widths_lightcurves}a) and SJI-1400 (Fig.\ \ref{fig:widths_lightcurves}b). In both passbands, the ribbons appear and begin to widen early in the flare; thereafter, the region grows much brighter as the ribbon fronts widen only marginally late in the flare. 

\begin{figure}
\centering
\includegraphics[width=9.5cm]{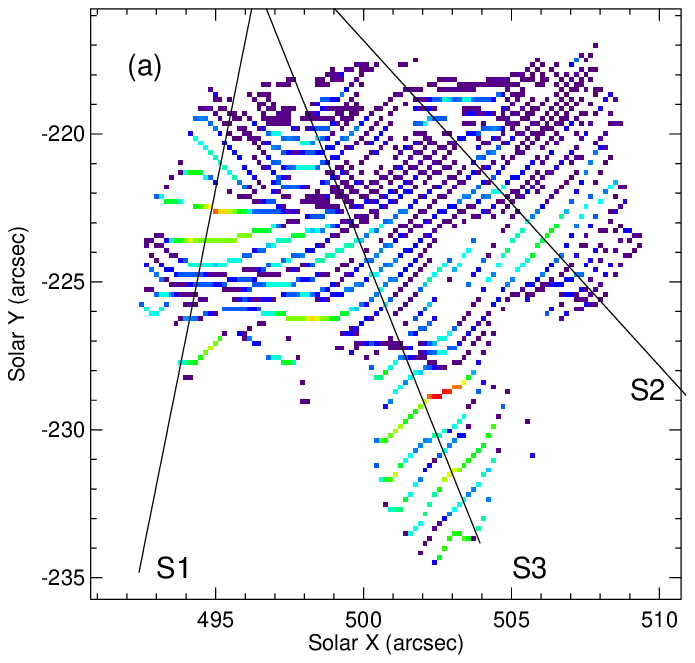}\includegraphics[width=7.5cm]{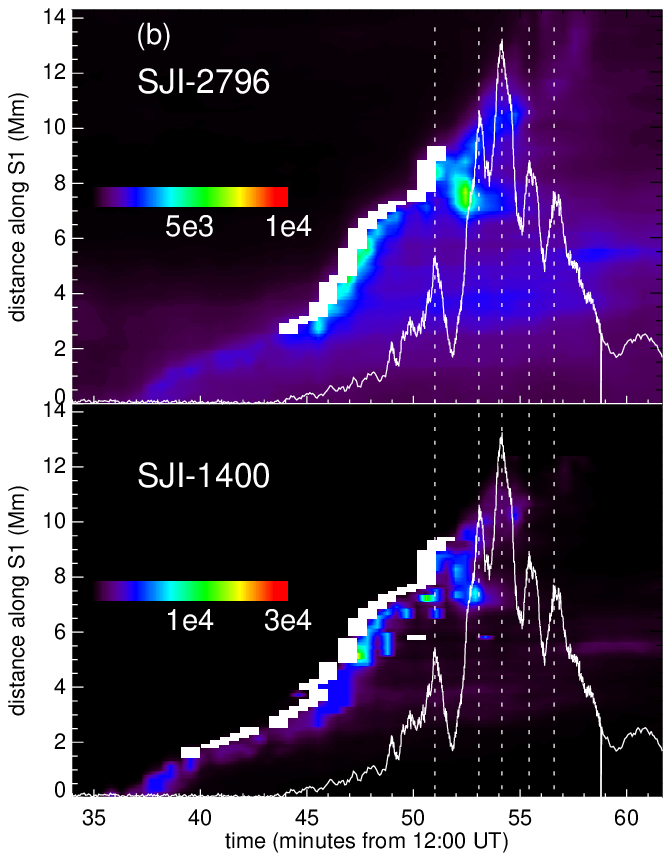}
\includegraphics[width=7.5cm]{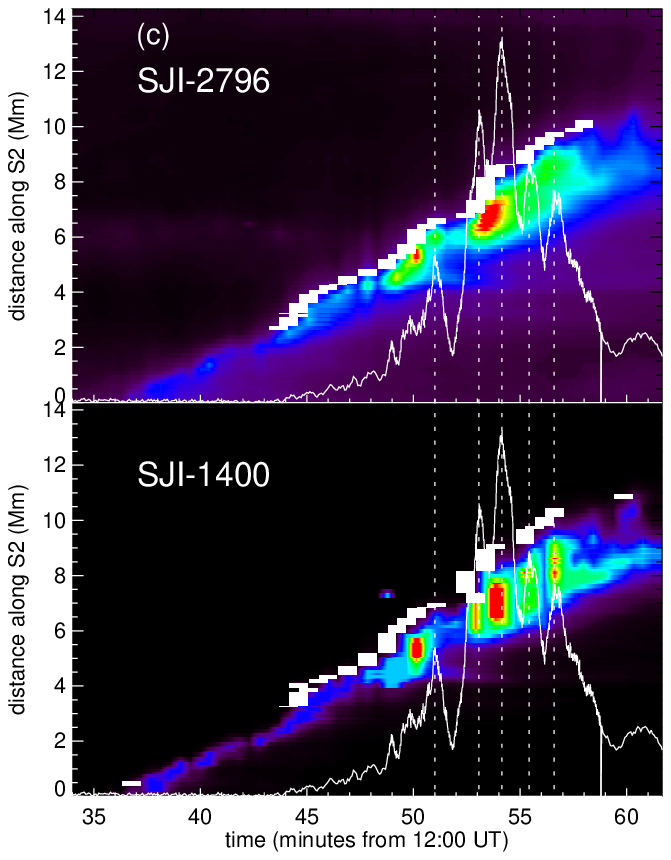}\includegraphics[width=7.5cm]{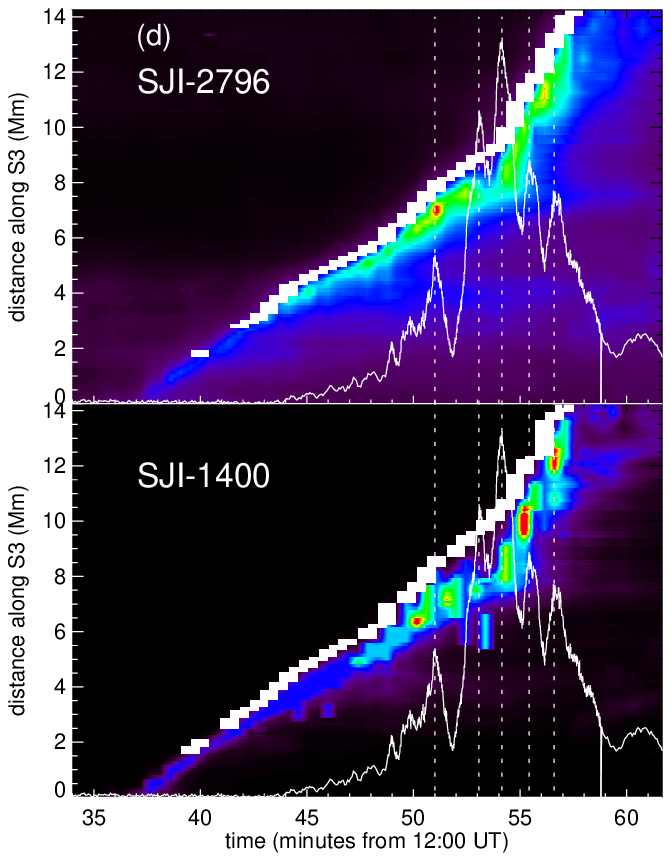}
\caption{Time-distance diagrams along three selected slits in ROI-1. (a) Positions of the slits S1-3 across different locations of the flare ribbon, shown on the SJI-2796 map with $N$ = 8. The field of the view is the same as that in Figure~\ref{fig:zebra2796} and Figure~\ref{fig:zebra1400}. (b)-(d) Time/distance diagrams of the ribbon brightness along each slit from north to south, derived from {\em IRIS} SJI-2976 (top) and SJI-1400 (bottom). The brightness is linearly scaled between 100 and 10000 DN for SJI-2796, and between 300 and 30000 DN s$^{-1}$ for SJI-1400. The height of the white rectangle bars denotes the width of the newly brightened ribbon fronts along the slit, measured with $N = 8$ for SJI-2976 and with $N = 100$ for SJI-1400. Also plotted is the HXR light curve at 24-51~keV obtained by {\em Fermi}/GBM (white); vertical dashed lines denote the HXR bursts as in Figure~\ref{fig:overview}e. 
\label{fig:tdplot}}
\end{figure}



A closer examination of local enhancements in the ribbon width and ribbon intensity substantiates the conclusions suggested by the mean ribbon-front width and total light curves. We construct  time-distance diagrams of the UV brightness and ribbon width in ROI-1 along three selected slits, S1-3, shown in Figure~\ref{fig:tdplot}a. {The slits are chosen so that each of them is roughly perpendicular to the parts of the ribbons it crosses, and each crosses a region with enhanced ribbon width, guided by the map of the spatio-temporal distribution of the UV ribbon width.} 
The ribbon brightness along each slit, as observed in both SJI-2796 and SJI-1400, is displayed as color shading in Figure~\ref{fig:tdplot}b-d. We also plot the measured ribbon-front widths along each slit as white rectangular bars: the height of each bar denotes the width of the newly brightened ribbon front at the given time and location along the slit. 
In agreement with the global measures shown in Figure~\ref{fig:widths_lightcurves}, the enhancements in the ribbon-front width consistently precede the peaks in ribbon intensity. The lead time is only 1-2 time frames in some cases, but it is several time frames in others, especially for the strongest enhancements in brightness (e.g., SJI-1400 along slit S2, Fig.\ \ref{fig:tdplot}c). 

We include in Figure~\ref{fig:tdplot} the HXR light curve from {\em Fermi}/GBM in the 24-51~keV band and vertical dashed lines noting the times of HXR bursts identified in Figure~\ref{fig:overview}e. The  strongest UV brightness enhancements (red) in the ROI-1 ribbon fronts coincide quite closely with peaks in the HXR light curve (white) while the width increases typically slightly precede the HXR peaks. Although they are not definitive, these associations are highly suggestive of connections between the UV ribbon brightness and width enhancements on the one hand, and the HXR bursts on the other. We explore this link further in \S\ref{subsec:hxr} below.


\subsection{Connection with Magnetic Flux Changes}
\label{subsec:magflux}

\begin{figure}[ht!]
\centering
\includegraphics[width=9cm]{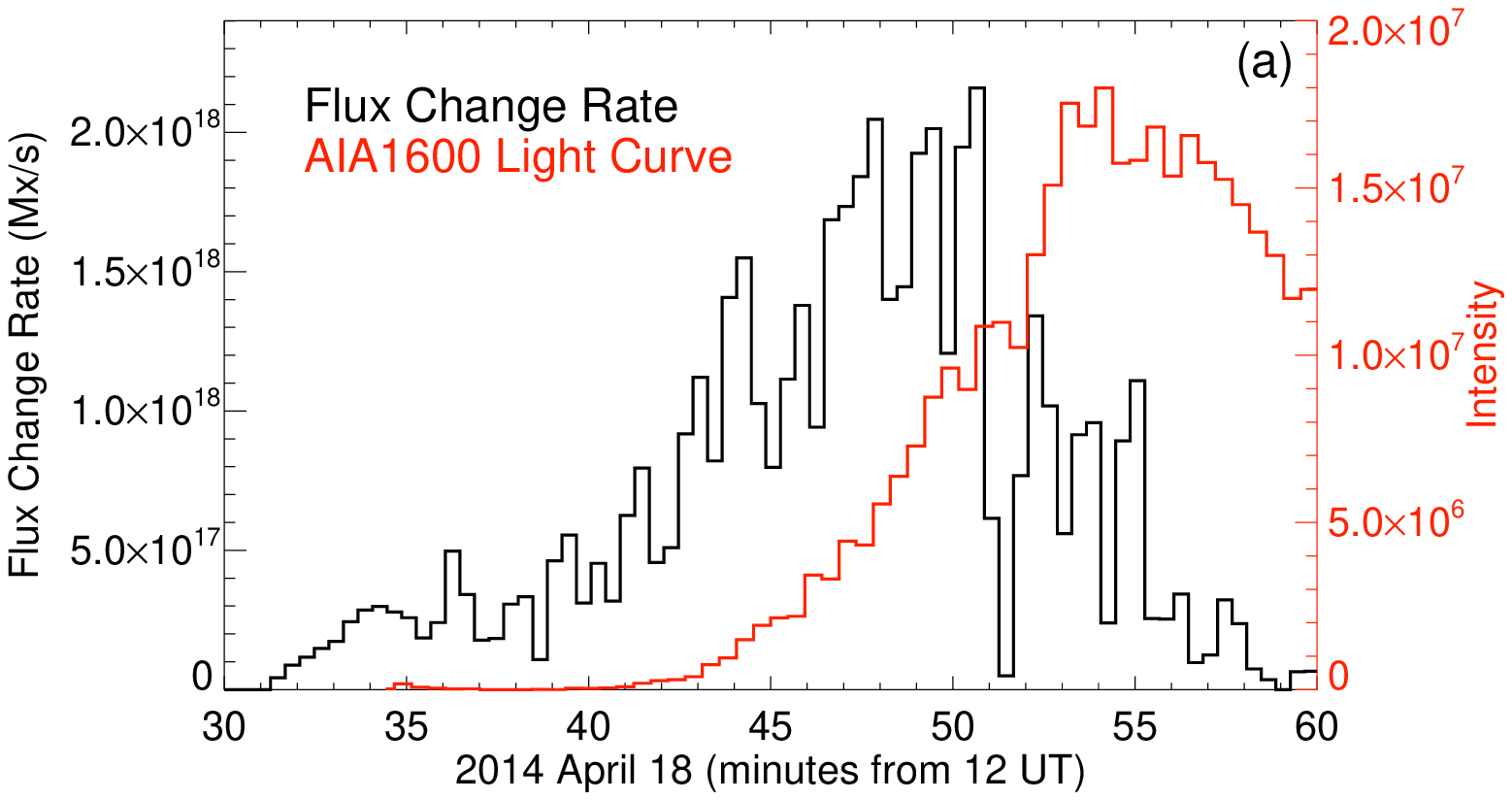}\includegraphics[width=9cm]{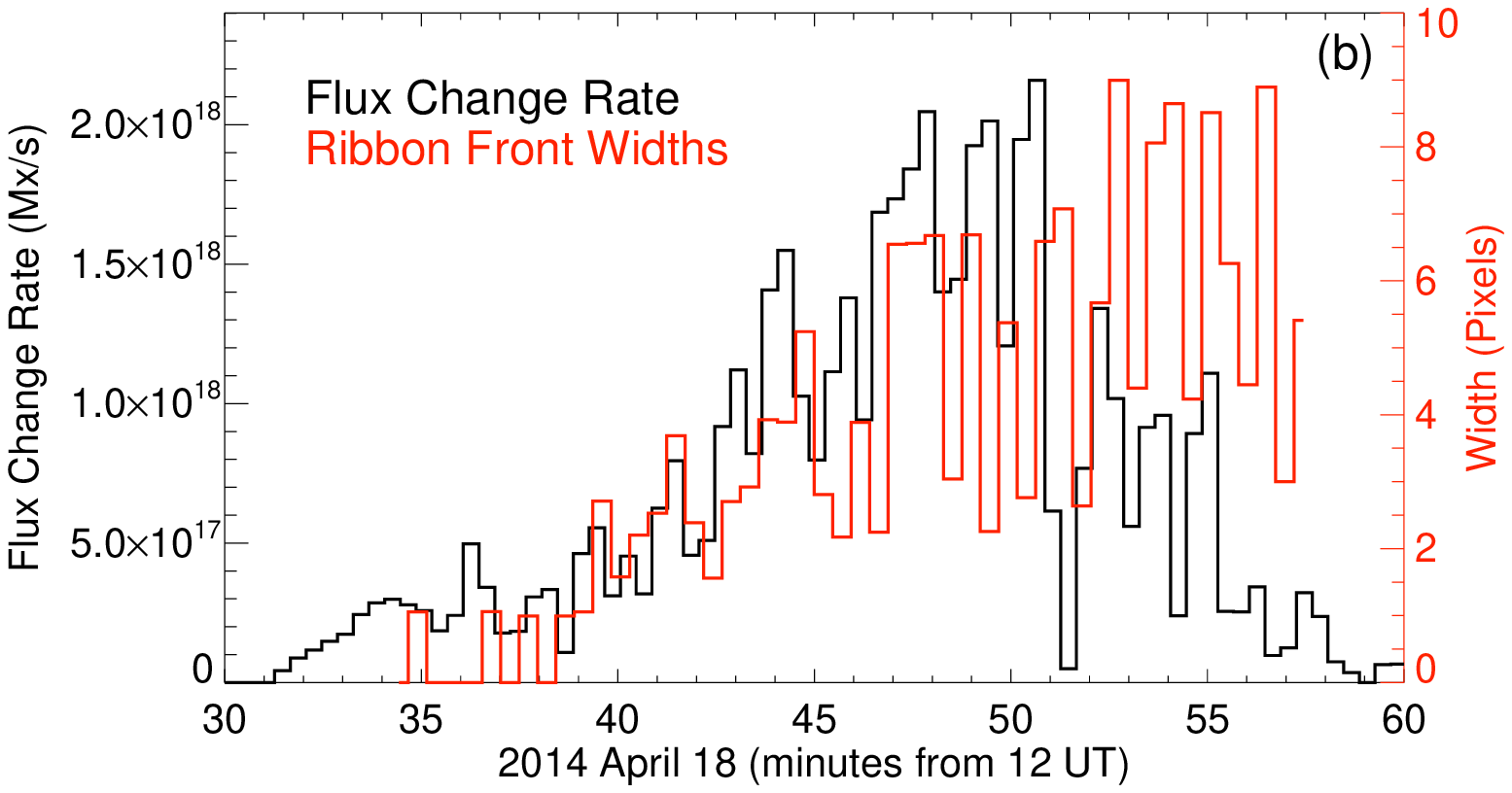}
\includegraphics[width=9cm]{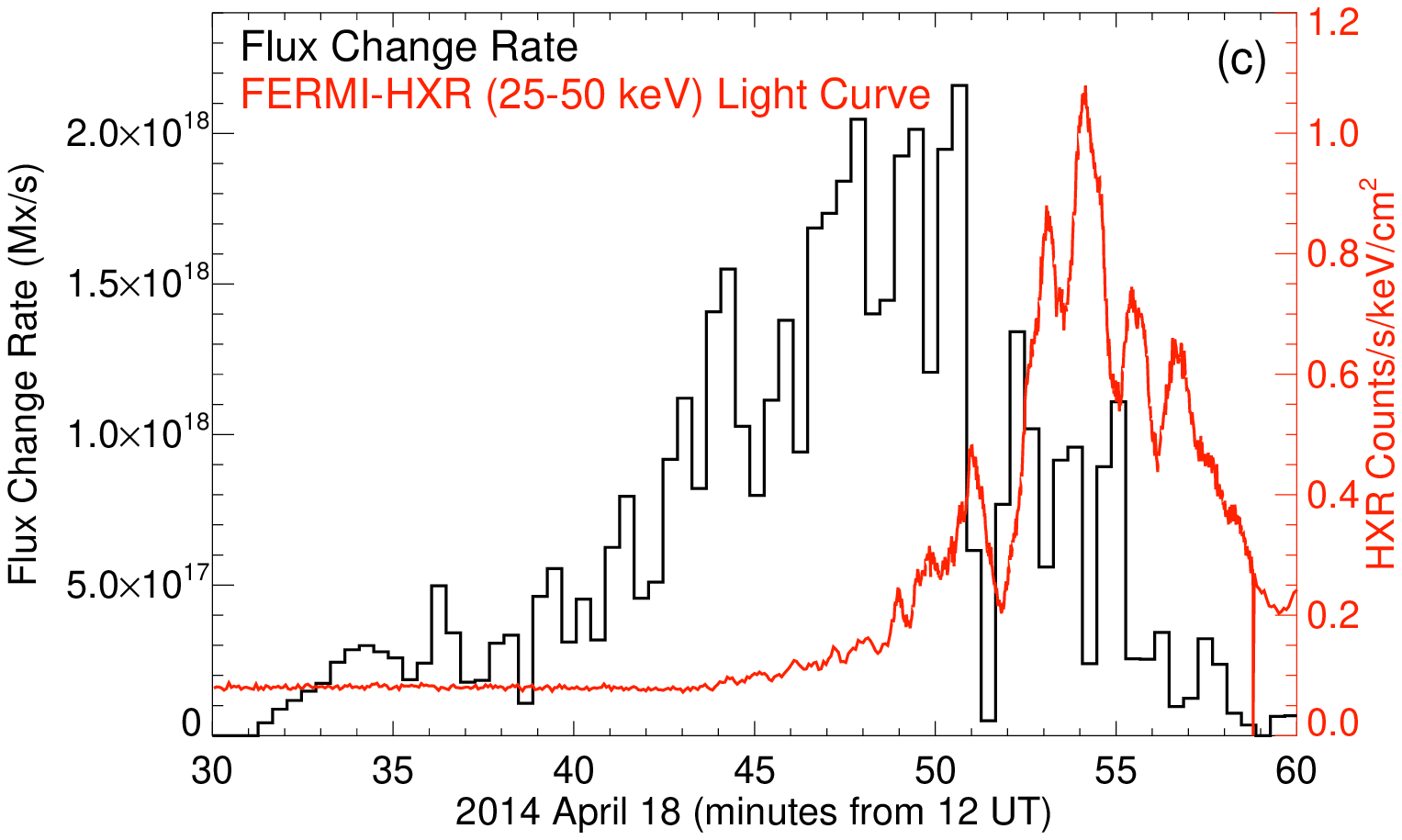}\includegraphics[width=9cm]{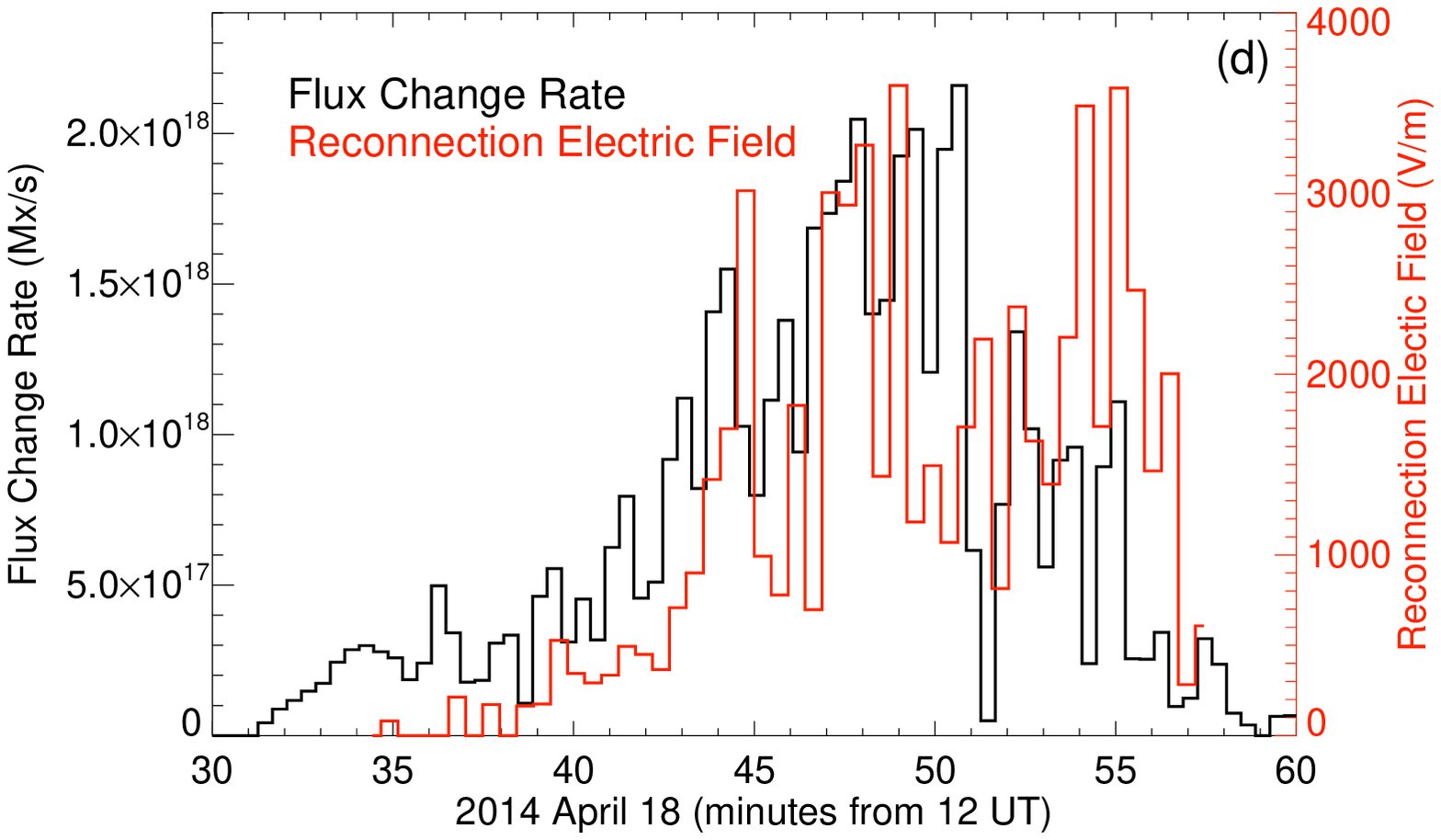}
\caption{The rate of the magnetic flux change in ROI-1, measured with the radial magnetic field $B_r$ from HMI and newly brightened ribbons areas derived from the AIA-1600 UV images, shown along with the region's (a) AIA-1600 light curve, (b) SJI-1400 top 5 ribbon front widths (for $N$ = 100), (c) FERMI-HXR light curve for 24-51 keV, and (d) the reconnection electric field inferred from the product of the SJI-1400 top-5 ribbon front widths (for $N$ = 100) and magnetic field divided by the cadence of the SJIs $\tau = 28$~s (see text).
}
\label{fig:fluxregions}
\end{figure}

We have seen in the preceding section that the UV ribbon fronts begin to form and widen early in the flare, significantly prior to the attainment of peak UV brightness in our region ROI-1. Here, we relate these features to the magnetic-flux changes that were measured in ROI-1 using the techniques described in \S\ref{sec:overview}. The change rate measured across the field of view encompassing the entire flare (Fig.\ \ref{fig:overview}b) was shown previously (Fig.\ \ref{fig:overview}f). In Figure~\ref{fig:fluxregions}, we plot the flux change rate (red curves) calculated from the AIA-1600 UV ribbons and the HMI magnetic data in ROI-1 alone. ROI-1 included most of the positive-polarity portion of the UV ribbons and essentially none of the negative-polarity portion. 
The flux change rises steeply and approximately linearly, with fluctuations, up to about time 12:51~UT, when it suddenly drops to about half its peak magnitude. After a short plateau phase, the flux change quickly decreases toward zero beginning at about 12:55~UT. 

In Figure~\ref{fig:fluxregions}, the rate of change of magnetic flux in ROI-1 is compared with the overall AIA-1600 light curve (Fig.\ \ref{fig:fluxregions}a), the SJI-1400 ribbon widths (Fig.\ \ref{fig:fluxregions}b), both integrated in ROI-1, and the spatially unresolved HXR (25-50 keV) light curve (Fig.\ \ref{fig:fluxregions}c). The comparison with the AIA-1600 light curve shows that the flux-change rate transitions (at about 12:51~UT) when the UV flare brightness in the region has reached only about 50\% of its peak value (which occurs at about 12:55~UT). This is reminiscent of the previously noted saturation of the ribbon-front widths (at about 12:51~UT) shown explicitly here in the second panel. The ribbon-front widths and the flux-change rates evolve in quite close coordination until about this time. Thereafter, the flare brightness continues to climb and the ribbon-front widths also increase slightly, whereas the flux-change rate drops by half and then decreases rapidly toward zero. These features indicate that the newly brightened pixels contributing to the ribbon fronts in the late flare phase increasingly are located in weak-field regions, so they make decreasing contributions to the flux-change rate. Evidence for this is clear in Figure~\ref{fig:zebra}a-b, which shows that the measured magnetic-field strengths are smaller where the UV ribbons form later, in the southwest portion of ROI-1. 

Finally, if we assume that the ribbon width $\delta$ reflects the velocity of newly reconnected field lines ``traveling" across the chromosphere in the direction perpendicular to the reconnection current sheet in the corona, we may also estimate the reconnection electric field $E = v_{in}B_{in} \approx \delta \langle B_r\rangle/\tau$ \citep{Forbes1984}. Here $v_{in}$ and $B_{in}$ are the inflow speed and inflow magnetic field strength in the corona, respectively, $\tau$ is the cadence of SJIs, and $\langle B_r\rangle $ is the average radial photospheric magnetic field underlying the ribbon fronts. Figure.\ \ref{fig:fluxregions}d shows the time evolution of $E$, estimated using the top 5 width measured with the SJI-1400 data in ROI-1, in comparison with the flux change rate measured with the AIA-1600 data. Before 12:50~UT, the two measurements roughly track each other; afterwards, the estimated $E$ first decreases, but then increases again. This result shows the different patterns of the time evolution between the {\em global} reconnection rate ($\dot{\psi}$), and the possibly maximum {\em local} reconnection rate ($E$), particularly after 12:50~UT.  

\subsection{Connection with HXR Emissions}
\label{subsec:hxr}
\begin{figure}[ht!]
\center
\includegraphics[height=6cm]{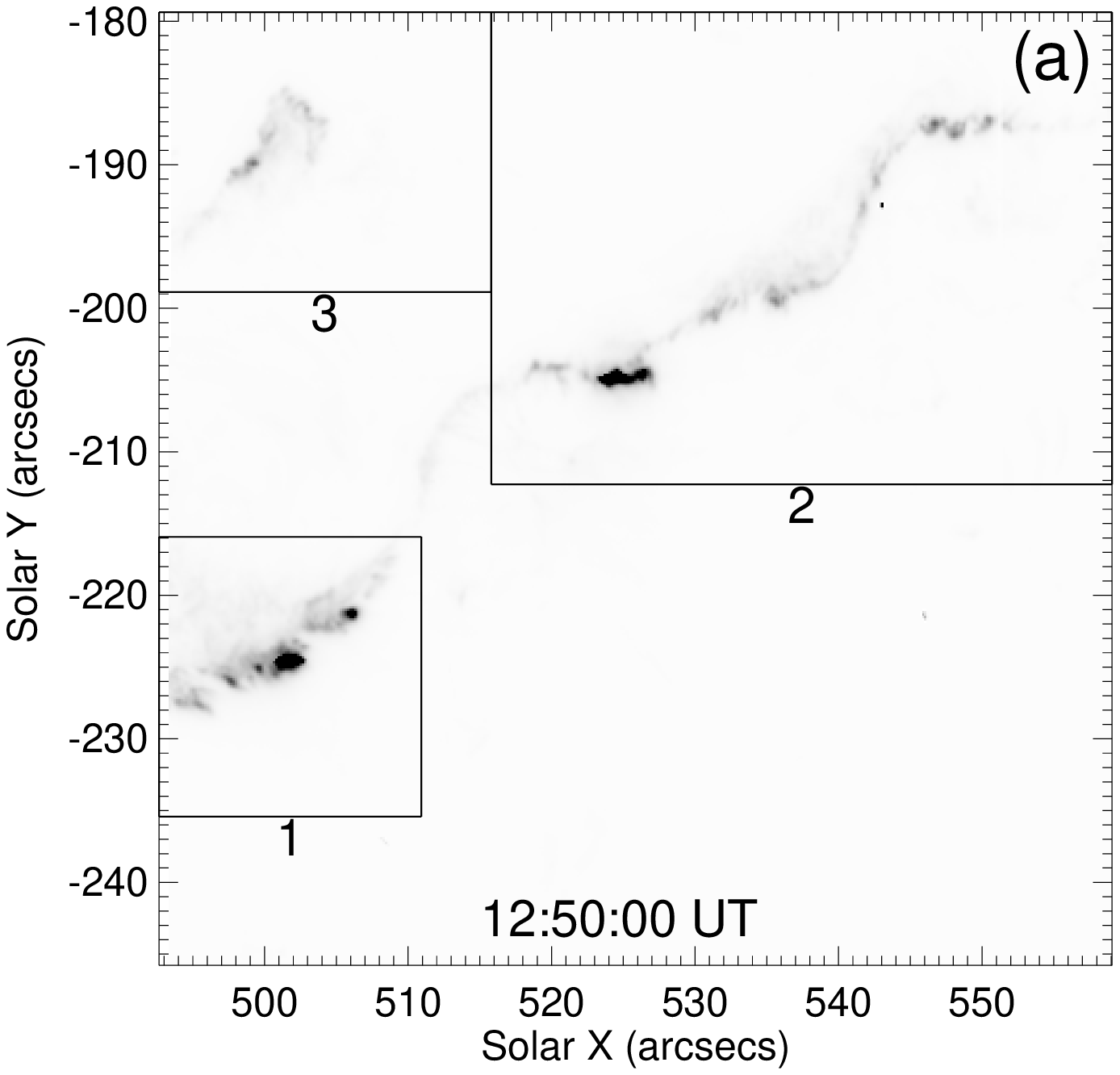}\includegraphics[height=6cm]{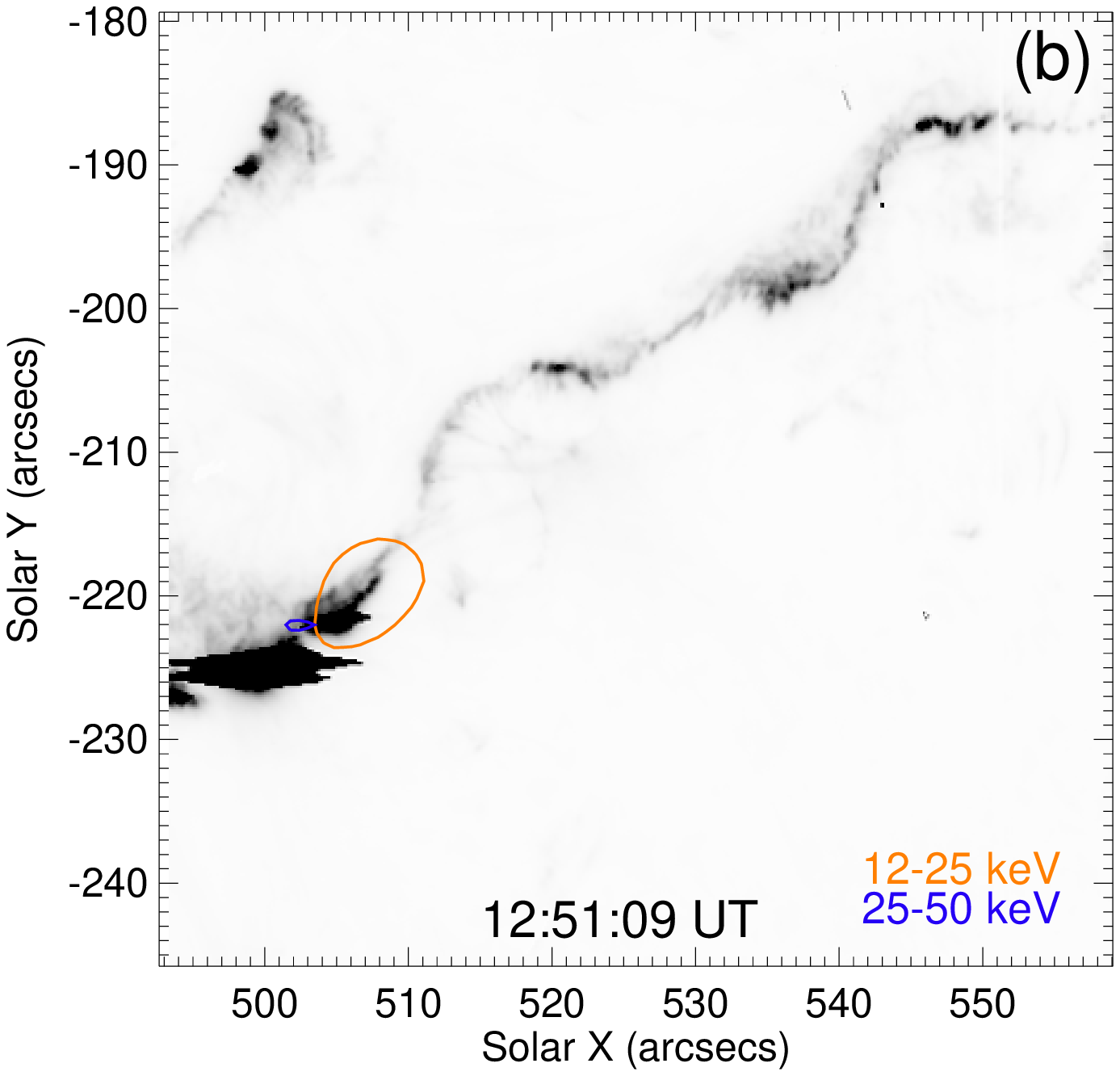}\includegraphics[height=6cm]{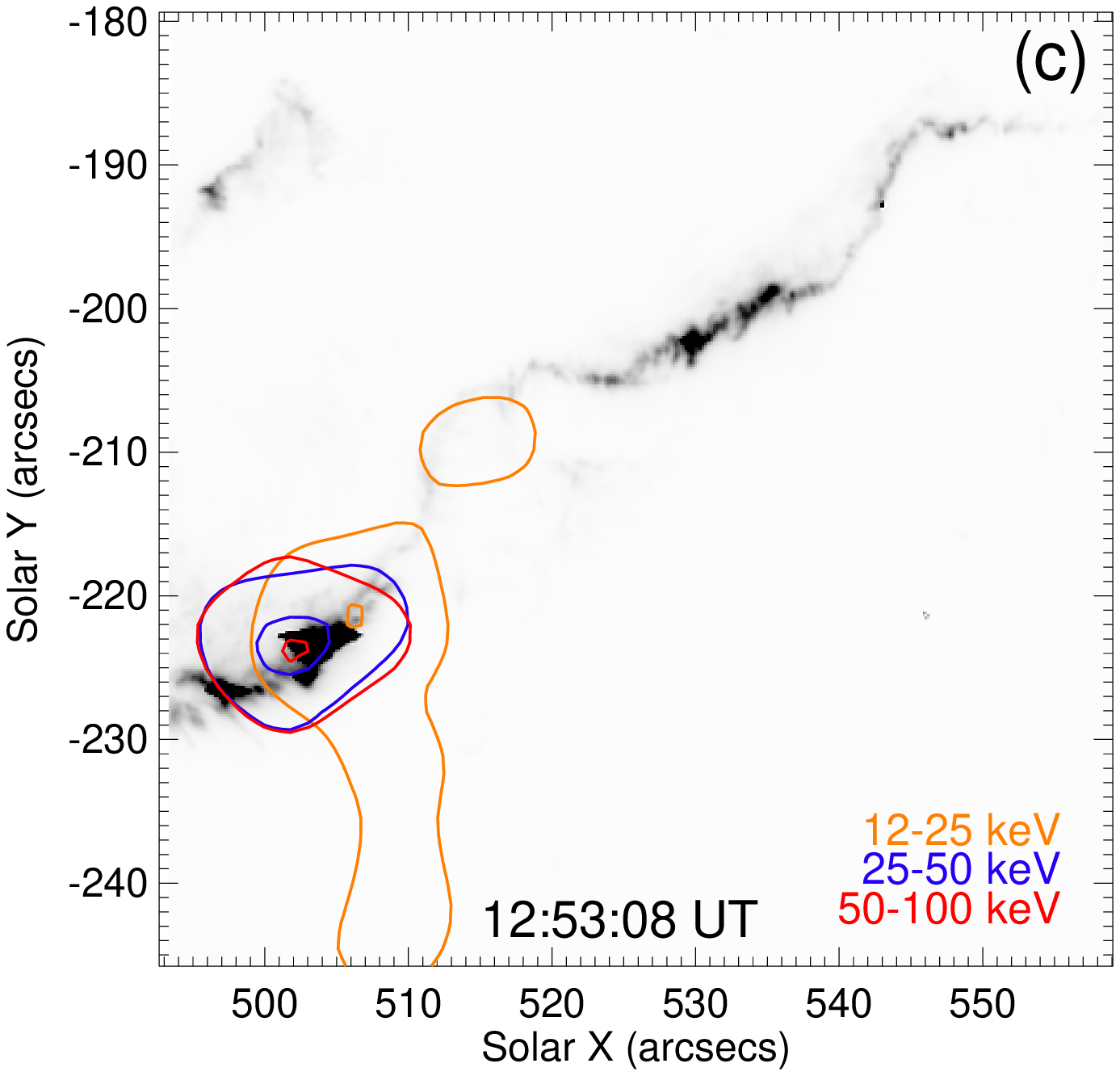}
\includegraphics[height=6cm]{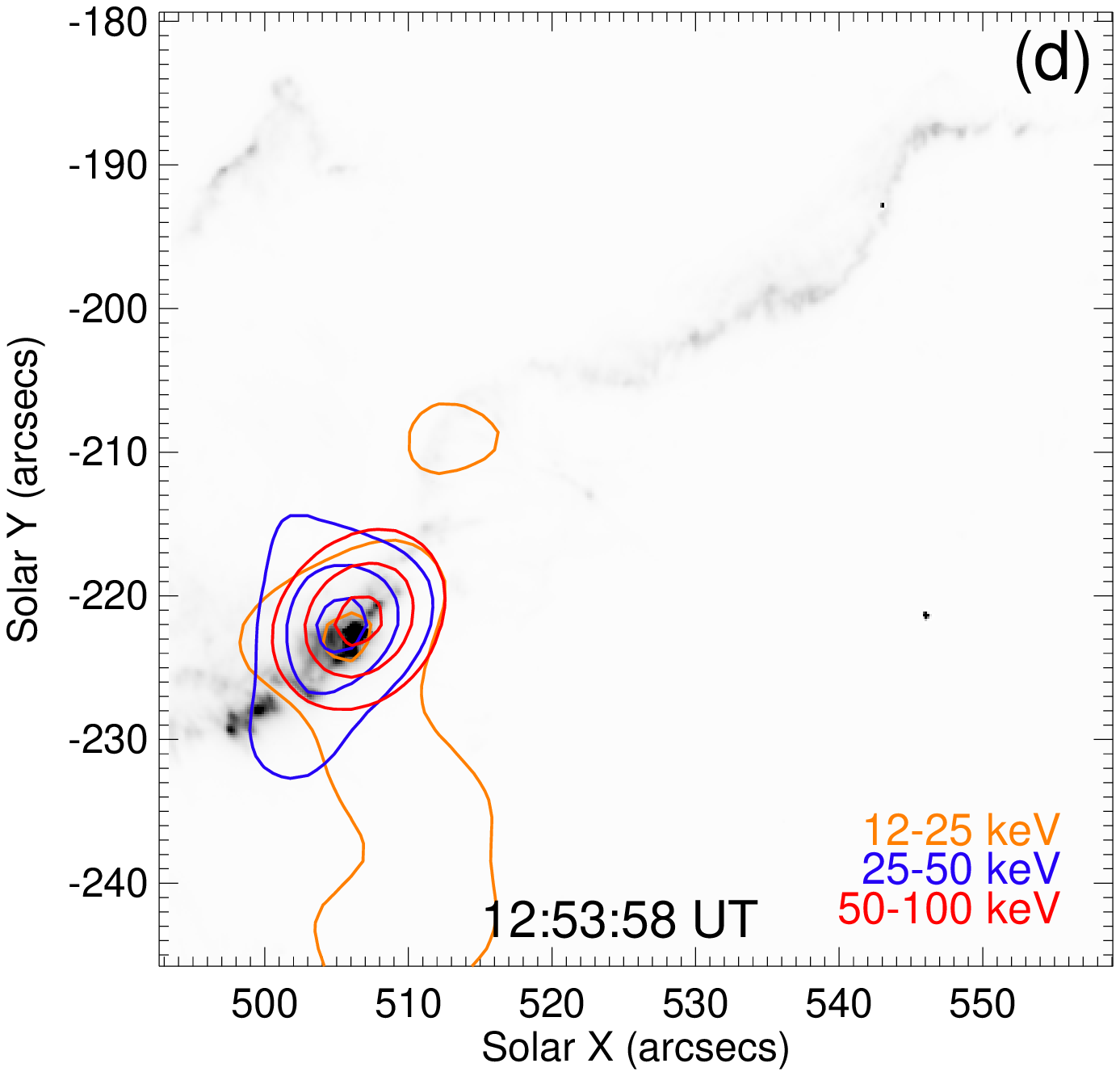}\includegraphics[height=6cm]{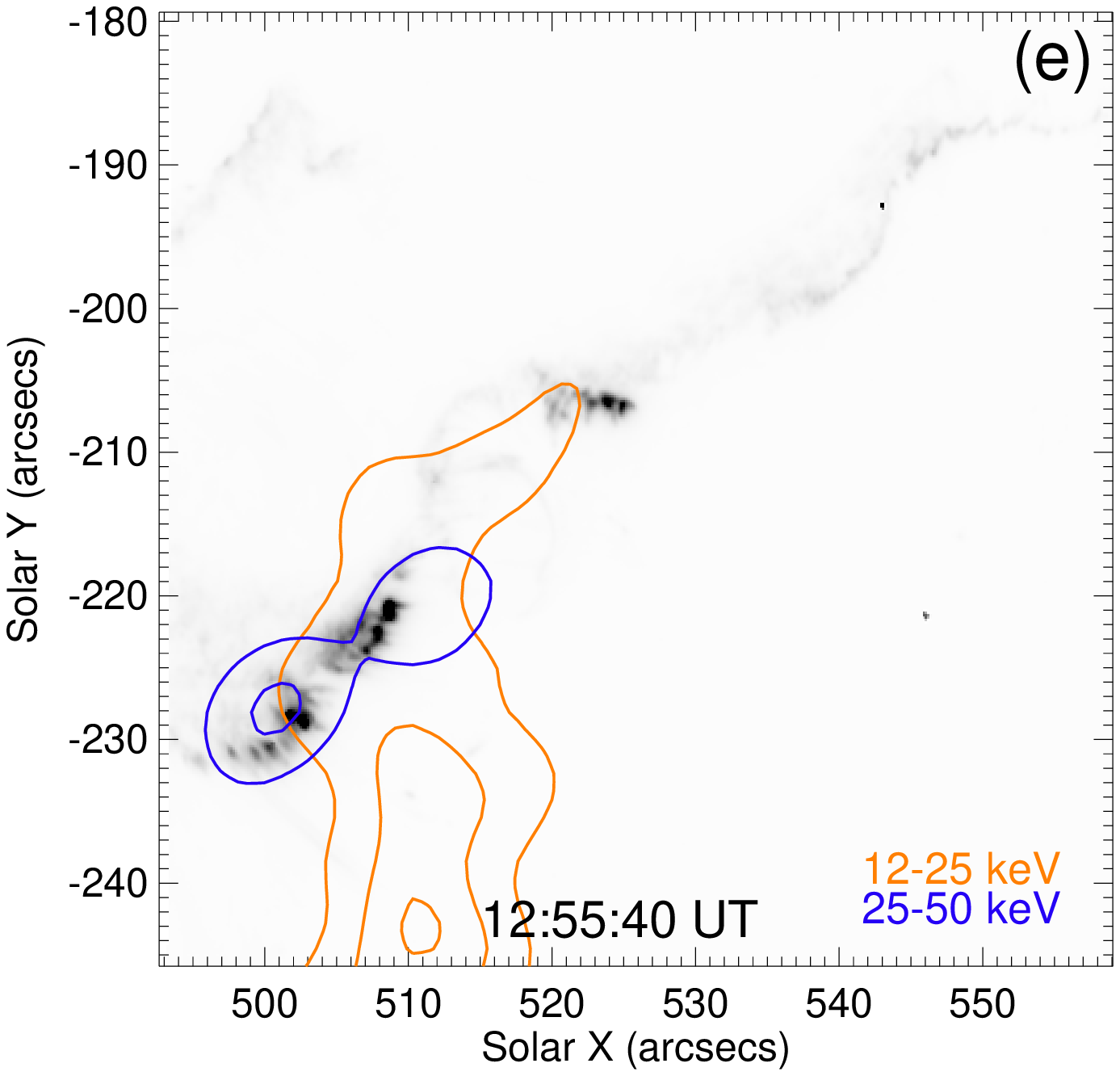}\includegraphics[height=6cm]{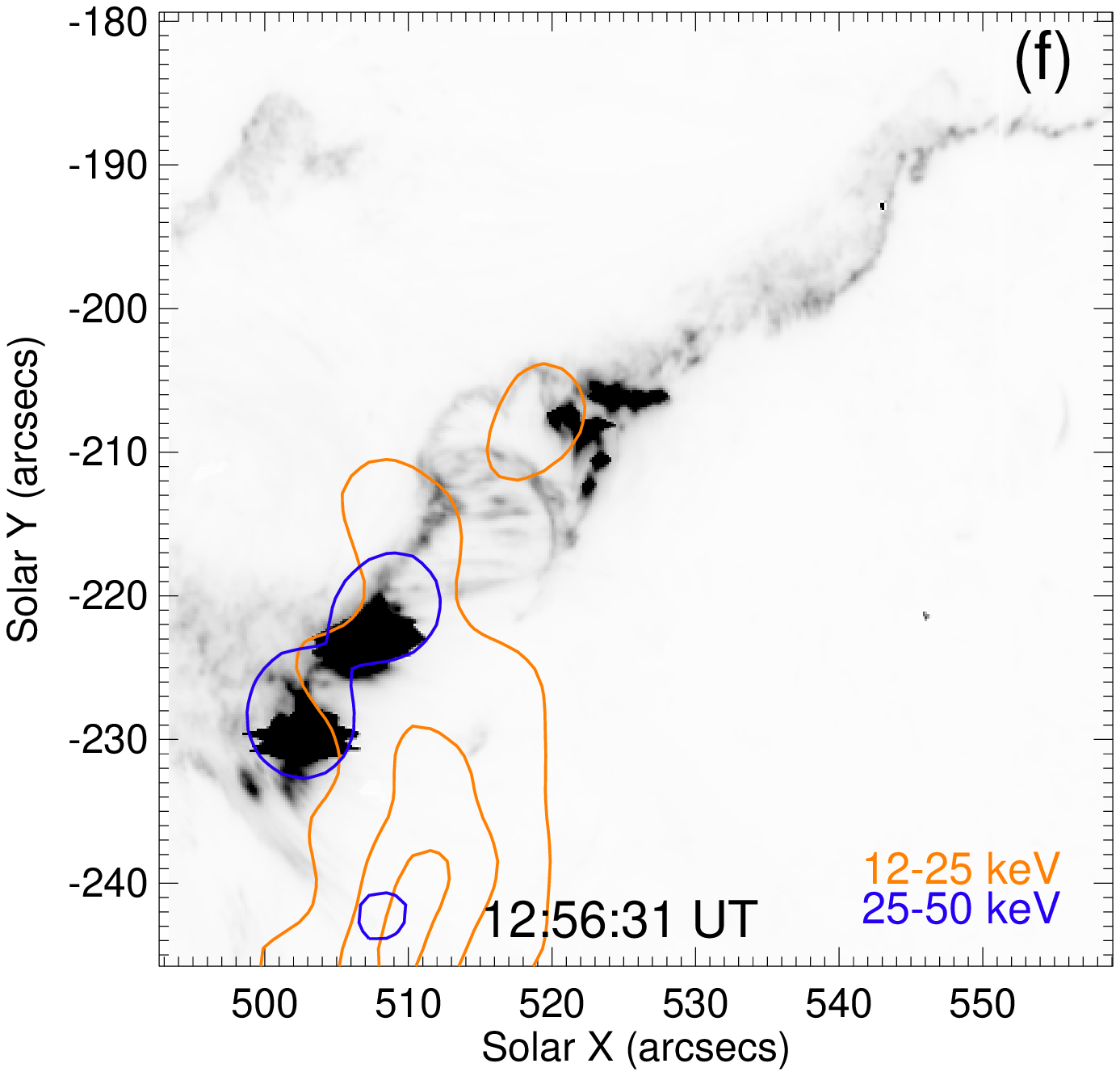}
\caption{(a) ROI-1, -2, and -3 superimposed on an {\em IRIS} image in SJI-1400. (b-f) {\em RHESSI} contours at energy levels 12-25 keV, 25-50 keV, and 50-100 keV (orange, blue, and red colors) superimposed on {\em IRIS} images in SJI-1400 of the flare ribbons. The contours for each energy level are taken at 30\%, 60\% and 90\% of the maximum intensity across the time interval from 12:50 to 12:56~UT, at the times of peaks in the {\em Fermi}/GBM integrated light curve at 24-51 keV. 
The heliographic coordinates shown represent the position of the flare region at 12:40:16~UT.}
\label{fig:rhessi}
\end{figure}

As shown in Figure~\ref{fig:overview} and discussed in previous subsections, the UV flare ribbons began to form at about 12:40~UT; however, the HXR emissions began to ramp up only just before 12:50~UT (Fig.\ \ref{fig:overview}e). At the latter time, {\em RHESSI} emerged from behind Earth and began to collect HXR data of the Sun from the flare already in progress.  Figure~\ref{fig:rhessi} shows the evolution of the flare as observed by both {\em IRIS} in SJI-1400 (gray scale) and by {\em RHESSI} (color contours)\footnote{HXR maps are obtained from the {\em RHESSI} image archive \url{ https://hesperia.gsfc.nasa.gov/rhessi\_extras/flare\_images/hsi\_flare\_image\_archive.html}. They are constructed by applying the CLEAN algorithm to data from detectors 3 to 9, and the integration time of each map varies from 48 to 64~s. Only maps with the signal-to-noise ratio larger than 5 are illustrated in this paper.} within the same field of view shown in Figure~\ref{fig:ribbon}c. The selected times chosen correspond approximately to the peak times of the HXR bursts in the {\em Fermi}/GBM light curve at 24-51 keV (Fig.\ \ref{fig:overview}e, in blue). 

Figure~\ref{fig:rhessi}a shows the {\em IRIS} image only at 12:50~UT, with ROIs 1-3 marked as in Figure~\ref{fig:ribbon}c for context when examining the remaining panels b-f. The panels thereafter (Fig.\ \ref{fig:rhessi}b-d) show clearly that the HXR emission at $>$25 keV is concentrated on the ribbon section in ROI-1 throughout the 5 bursts. 
The {\em RHESSI} contours at lower energies at these times extend to the south and east from ROI-1, toward the negative-polarity portions of the magnetic flux density and the UV ribbons, as can be seen in the larger AIA+HMI field of view in Figure~\ref{fig:overview}a and b. Due to the on-disk position of the flaring region, it is likely that $<$ 25 keV HXR contours also reflect contributions from loop-top sources in the overlying coronal loop connecting the positive and negative ribbons.  
At the last times shown (Fig.\ \ref{fig:rhessi}e-f), the expanding contours at 12-25 keV encroach onto the southeast end of ROI-2 at about position (525$''$,$-$205$''$). Because the expansions are not centered on the dark kernel of the SJI-1400 ribbon, it seems less likely that these emissions are footpoint sources within ROI-2, although we cannot rule this out entirely. \citet{Brosius2016} similarly concluded from their own {\em RHESSI} maps that the $>25$~keV HXR emissions observed during this flare were concentrated principally within, or immediately adjacent to, our ROI-1. 

\begin{figure}[ht!]
\centering
\includegraphics[width=9cm]{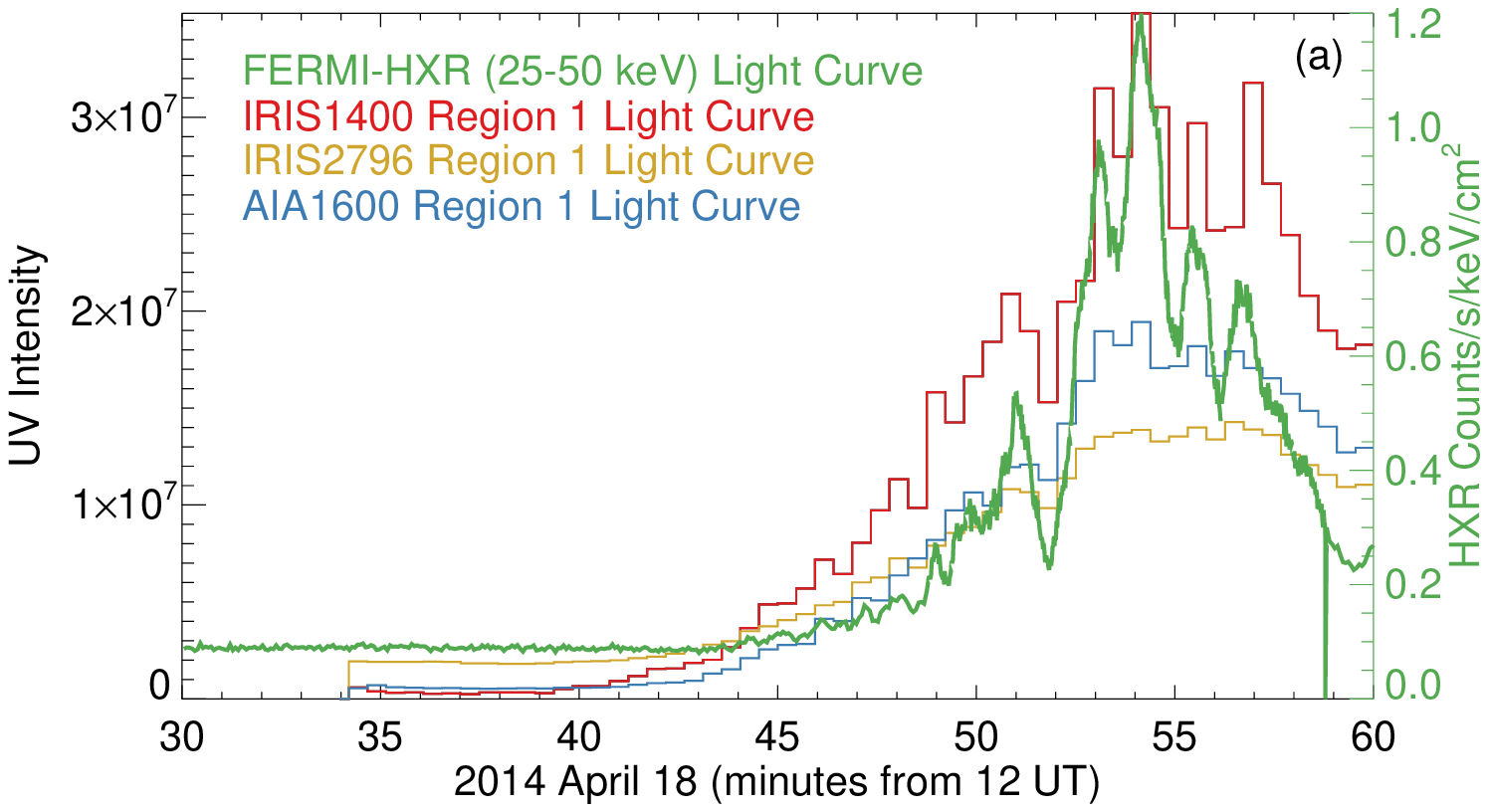}
\includegraphics[width=9cm]{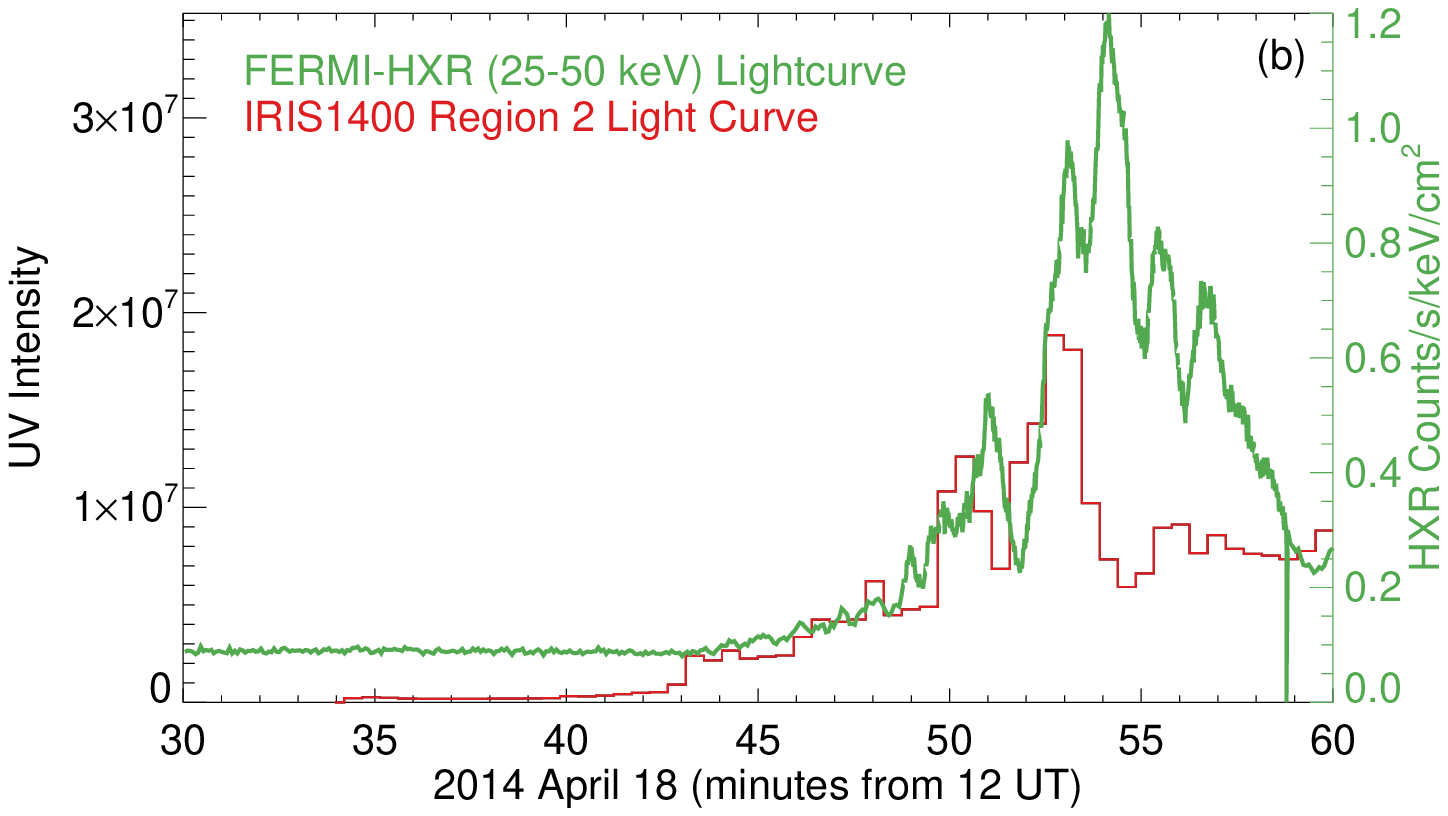}\includegraphics[width=9cm]{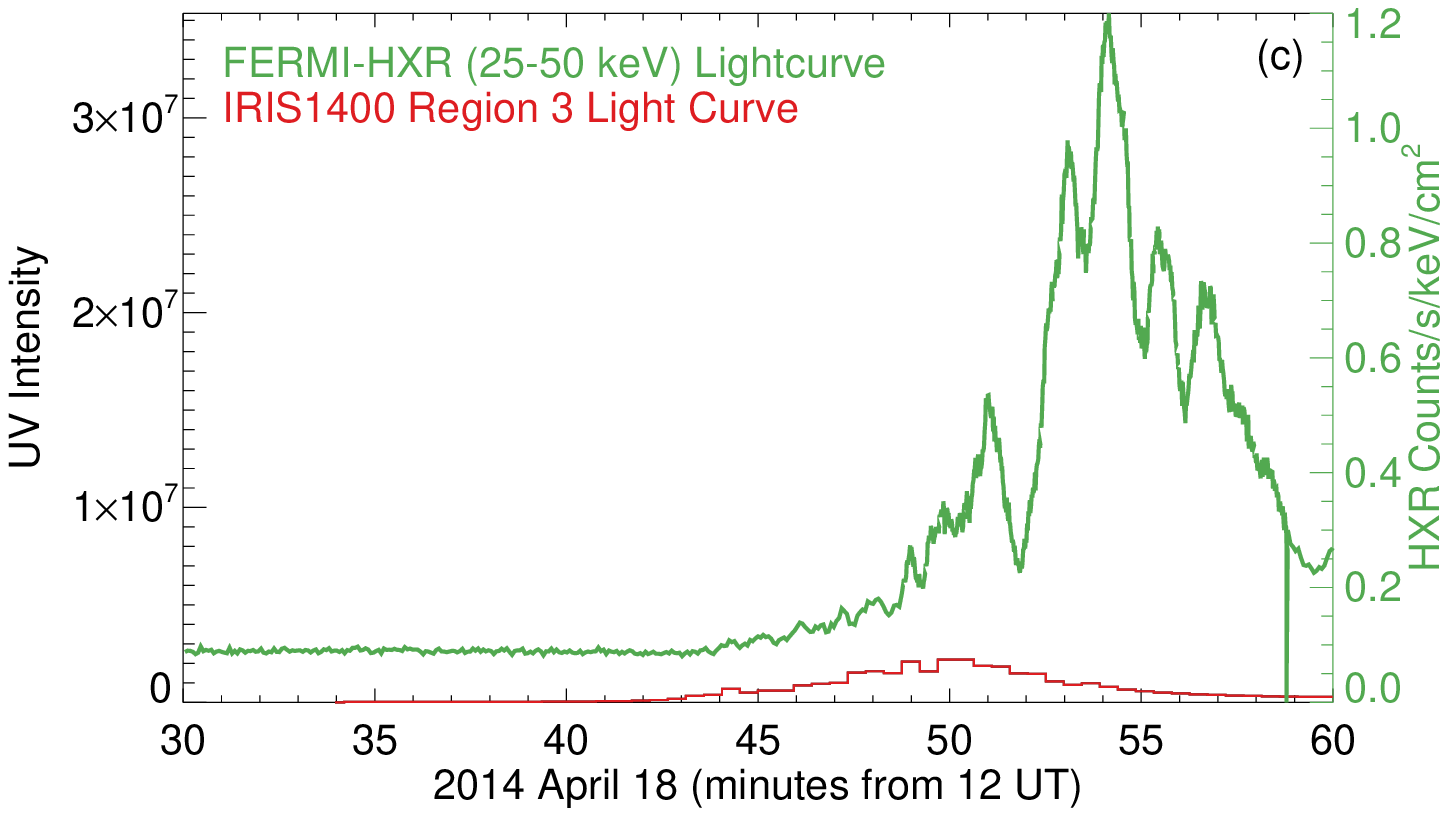}
\caption{{\em Fermi}/GBM HXR light curve in the 24-51 keV band (green) plus: (a) ROI-1 UV light curves in SJI-1400 (red), SJI-2796 (yellow), and AIA-1600 (blue); (b) ROI-2 SJI-1400 light curve (red); and (c) ROI-3 SJI-1400 light curve (red). 
\label{fig:hxr}}
\end{figure}

Additional insight into the progress of the flare is given by Figure~\ref{fig:hxr}, which shows in panel (a) the three UV light curves in ROI-1 (SJI-1400, SJI-2796, and AIA-1600) along with the full-disk 24-51 keV light curve from {\em Fermi}/GBM. Each of the four curves has a broad maximum centered roughly at 12:55~UT, albeit with substantial fluctuations, bolstering our conclusion (and that of \citet{Brosius2016}) that the HXR emissions appear to originate primarily in ROI-1. The curves also show that the UV emissions increase, gently at first, beginning at about 12:40~UT. In contrast, the HXR emissions remain flat for an additional few minutes, beginning their own initially gentle increase at about 12:45~UT. Strong spikes in the HXR data occur contemporaneously with spikes in the UV data, especially SJI-1400, although there is no consistent one-to-one correspondence between the curves. 



We also show in Figure~\ref{fig:hxr} the SJI-1400 light curves from regions (b) ROI-2 and (c) ROI-3 along with the HXR 24-51 keV light curve. UV intensity peaks at about 12:51~UT in both regions, and at about 12:53~UT in ROI-2, align well with HXR peaks at those times, albeit with significantly lower amplitudes in ROI-2 and a very low amplitude in ROI-3 compared with the corresponding SJI-1400 peaks in ROI-1. On the other hand, there are no UV peaks at the times of the remaining HXR peaks in ROI-2 and ROI-3. These comparisons further comfirm that, in the positive magnetic fields, $>25$~keV HXRs are rather localized and produced primarily in ROI-1. 

\begin{figure}
\includegraphics[width=18cm]{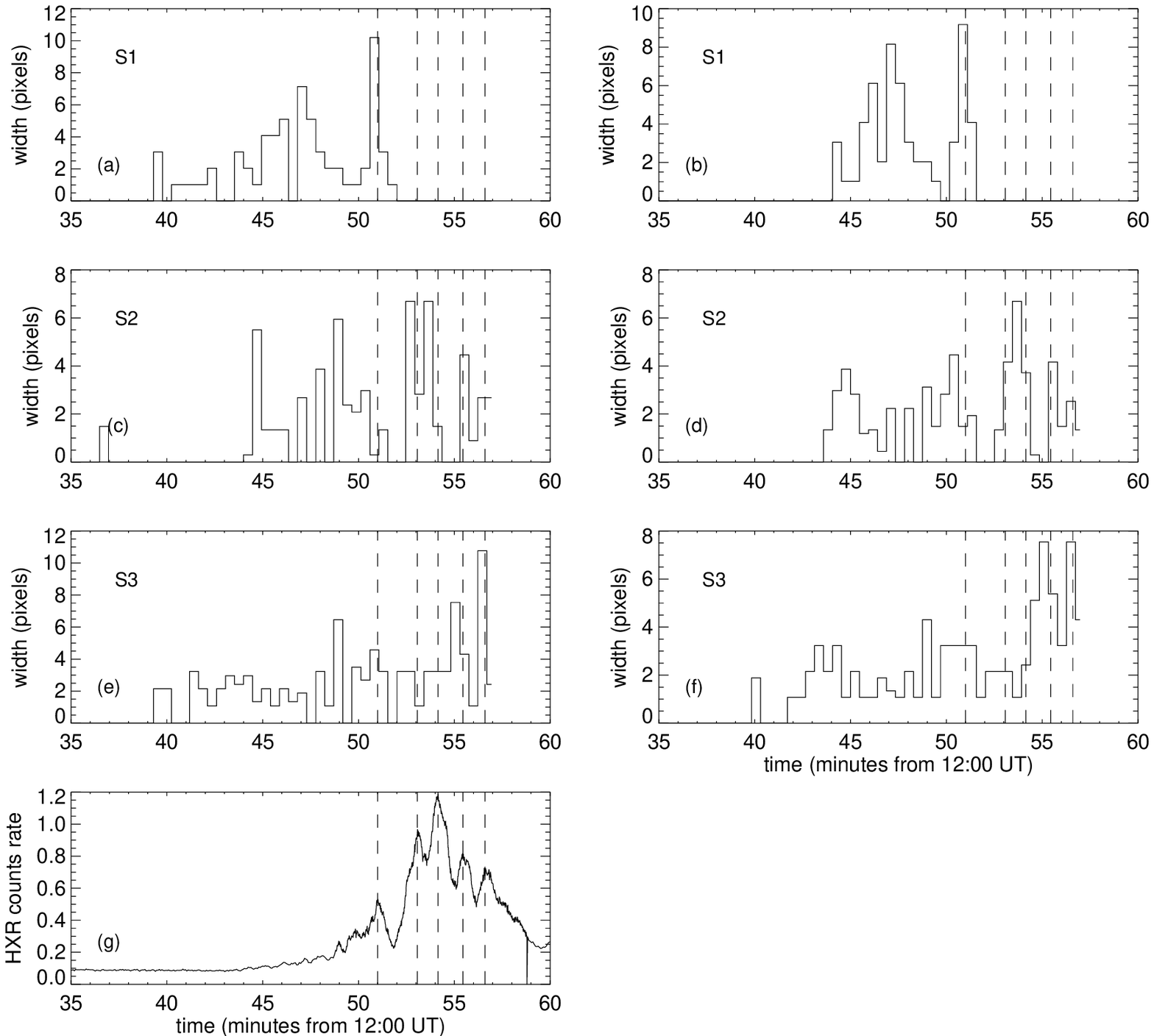}
\caption{(a-f) Line plots of the peak UV ribbon-front width 
along slits S1 (a,b), S2 (c,d), and S3 (e,f) shown in Figure~\ref{fig:tdplot}, using the SJI-1400 (a,c,e) and SJI-2796 (b,d,f) data sets. (g) Line plots of {\em Fermi}/GBM 24-51 keV HXR data; dashed lines mark the principal HXR peaks and are included in (a-f). } 
\label{fig:hxreuv}
\end{figure}

Additional suggestive, although not definitive, evidence for an association between the local UV ribbon-front widenings and the HXR emissions in ROI-1 is presented in Figure~\ref{fig:hxreuv}. There, we replot the ribbon-front width data (black) previously shown in Figure~\ref{fig:tdplot} along the slits S1 (Fig.\ \ref{fig:hxreuv}a,b), S2 (Fig.\ \ref{fig:hxreuv}c,d), and S3 (Fig.\ \ref{fig:hxreuv}e,f) derived from the SJI-1400 (a,c,e) and SJI-2796 (b,d,f) data. 
Marked with vertical dashed lines are the times of peaks in the {\em Fermi}/GBM 24-51 keV light curve, which is shown separately (g,h). {The key feature to be noted here is that essentially all peaks in the HXR data are immediately preceded or accompanied by peaks in the UV ribbon-front width data from at least two measurements.}
It seems highly implausible that all of these close temporal superpositions of UV ribbon-widening events and HXR bursts are coincidental. On physical grounds, as argued in the Discussion below, there is ample reason to expect that local footpoint sources of hard X-rays should also be sources of UV brightness enhancements in the flare ribbons at the chromosphere. The observational evidence we have presented in Figure~\ref{fig:hxreuv} provides promising support for this expectation. However, high-resolution, direct X-ray imaging of the Sun would be required to definitively confirm or refute this connection.

\section{Discussion}
\label{sec:discussion}

In this study, we present some of the most detailed observations reported of the evolution of flare ribbon structure using {\em IRIS} slit-jaw images with high spatial resolution ($<$150~km pixel scale) and moderate time cadence ($\approx$30~s), and we examine its relationship with flare UV and HXR emissions. We find that, whereas the UV ribbon fronts exhibit a globally laminar and {quasi-1D} structure, the enhancements of the ribbon front width are highly structured in both space and time, indicating that the process forming the ribbons is highly nonuniform and intermittent. We also find that the flare evolution in this event consists of two phases, each lasting about 10~minutes. The evolution of the ribbon-front widths, flux-change rate, and UV and HXR emissions follows different patterns in these two phases. Finally, observations in multiple wavelengths suggest that late episodes of locally enhanced UV ribbon-widening and brightening are both co-spatial and co-temporal with unresolved bursts of hard X-rays.

There are several major implications of these observations for theoretical understanding of flare reconnection.
The first is that even for this event, which has geometrically complex ribbons implying a highly warped current sheet, the flare reconnection is quasi-2D in nature. This can be seen in the ribbon-front plots of Figures~{\ref{fig:zebra2796}} and {\ref{fig:zebra1400}}. Note that, to a large extent, the fronts appear as continuous line segments that form systematically outwards from the PIL. This is the signature of reconnection at a long-lived magnetic X line \citep[e.g.][]{Sturrock1966}. The observations do not provide strong evidence for a fractal evolution of the reconnected flux, as may be expected from turbulent volumetric reconnection. Although turbulent regions may develop inside the reconnection layers, which cannot be resolved by the present observations, the current sheet does not appear to be dominated by volumetric turbulence in which there is little coherence to the reconnection and energy release. {Recent high-resolution, 3D MHD simulations of an eruptive flare \citep[][in preparation]{dahlin21a} exhibit laminar patterns of reconnection similar to those observed.}

On the other hand, there is considerable structure in the evolving ribbon fronts. Reconnection appears to be bursty {in both time and space}, which is evident from the spatio-temporal variations of the ribbon-front widths (Fig.~\ref{fig:zebra}). If the ribbon width reflects the apparent motion of reconnecting field lines perpendicular to the direction of the macroscopic current sheet, for the 28-s cadence of the images, the maximum speed of apparent motion is about 40 km~s$^{-1}$ at the locations of greatest width. This is consistent with our time-distance diagrams along slits crossing various parts of the ribbon, which show increased speeds of ribbon spread up to 30 km~s$^{-1}$ (Fig.~\ref{fig:tdplot}) at the times and locations of large ribbon widths. These are among the highest speeds of perpendicular ribbon motion reported in the flare literature \citep[][and references therein]{Hinterreiter2018}. 
The locally enhanced ribbon width, and implied speed, is at least 3 to 4 times the {average measured in this flare (and note that the lower bound of the measurements is limited by the instrument resolution)}. It is unlikely that such local enhancement be due simply to noise in the system. The origin of the ribbon-width structure is unclear. It may be related to the concentrations of magnetic flux at the photosphere. However, the magnetic flux in the low-$\beta$ corona, where reconnection takes place, is certain to be much more smoothly distributed. Flare ribbons at the chromosphere are also observed to have typically much smoother structure than the flux distribution at the photosphere. Studies such as that by \citet{Fletcher2004}, but using high-resolution observations of both the photospheric and chromospheric magnetic-flux distributions, are needed to determine whether these distributions play any role in the variability of the reconnection. {{Furthermore, the response of the non-uniform and dynamic chromosphere to energy deposition from the corona may affect the brightness of the flare ribbons \citep[e.g.][]{Polito2018b, Cheung2021}, an effect that could be examined in future work. 
We suspect, however, that the bursty behavior observed in this study is primarily due to the reconnection process itself, perhaps related to the formation and ejection of magnetic islands within the coronal current sheet. }

{A possible alternative to the interpretation of the ribbon width increases simply as localized increases in the rate of magnetic reconnection is the possibility that the width increases are associated with the local broadening of the coronal current sheet. Such broadening is expected during reconnection in a 3D system where reconnection is not localized in a single current sheet \citep{daughton11a,dahlin15a,Huang16a}. Thus, although the observations are consistent with increased reconnection rate, as discussed previously, 
they do not preclude modest broadening of the current layer as documented in recent 3D flare simulations \citep{dahlin21a}. }

Regarding the flare energetics, observations suggest that the strong UV emissions in ROI-1 during the late phase are produced by non-thermal electrons, which also produce HXRs through thick-target bremsstrahlung \citep{Brown1971, Cheng1988, Warren2001, Qiu2010, Cheng2012}. The strong spatio-temporal connection between HXRs and the UV ribbon-width enhancements observed in ROI-1 during the late phase of this event seems consistent with several past studies, which indicated that non-thermal HXR or microwave emissions are, to varying degrees, temporally or spatially related to an enhanced reconnection electric field in some flares \citep{Qiu2004, Lee2006, Miklenic2007, Temmer2007}. On the other hand, there appears to be a small time lag, of order 30-120~s, between the enhancement of the ribbon-front widths and the peak UV/HXR bursts. Prior studies have reported similar time delays (as low as 20~s) of the peak UV or HXR emission with respect to the flux-change rate \citep{Miklenic2007}, lightly enhanced thermal emission \citep[e.g.][]{Hudson2021}, or initial chromospheric dynamic signatures due to deposition of reconnection-released energy \citep{Falchi1997, Czaykowska1999, Tian2015, Xu2016, Jeffrey2018, Panos2018, Graham2020, Kerr2021}. All of these observations seem to suggest that there is a finite time scale associated with the acceleration of non-thermal electrons, and/or an effective confinement of these electrons in the corona prior to their release to the chromosphere. 

We note that the onset of reconnection is determined, in all cases, by tracking the initial brightening of UV ribbons.
The implicit assumption is that this brightening indicates an initial energy release that travels down to the chromospheric footpoints fairly quickly, so that the observed ribbon evolution is due to the evolution of the flare reconnection rather than to some characteristic of the propagation process. This assumption seems reasonable: both an electron heat flux and an Alfv\'{e}n wave flux are likely to propagate at above 1,000 km s$^{-1}$, hence the flux would reach the chromosphere in only 10 s or less assuming a 10 Mm height of the X line. 

Based on this argument, we speculate that the time scale required for the UV ribbon to brighten from initial appearance to maximum brightness would reflect the time scale for continued energy release after reconnection onset. The observed time lag of 30-120 s is consistent with the standard model and simulation results for reconnection in a flare-like current sheet \citep[e.g.,][]{karpen12a}. When a small bundle of flux first reconnects at the X line, a small amount of energy is released associated with the flux breaking at the diffusion region. The bulk of the energy is released when the reconnected flux accelerates down through the current sheet as part of the outflow jets, eventually joining the flare arcade. The speeds that we obtain for the inflow and outflow seem consistent with the standard theory. For the observed reconnection bursts, we deduced speeds up to 40 km~s$^{-1}$ at the chromosphere. Because the magnetic flux is likely to expand somewhat as it extends upward from the chromospheric footpoints to the current sheet, the inflow speed may be expected to be about 100~km~s$^{-1}$; this agrees with the almost universally found reconnection rate of order 10\% \citep[e.g.][]{Birn2001, daughton14a, nakamura2018, Burch2020, Chen2020} for a 1,000 km s$^{-1}$ Alfv\'en speed. Assuming {a flare-loop height} of 10~Mm, together with the observed UV brightening time, implies an outflow speed of 100--500 km s$^{-1}$, which also agrees with both simulations \citep{guidoni16a} and coronal observations \citep[e.g.][]{Savage2011}. It appears, therefore, that these {\em IRIS} observations indeed are revealing to us the detailed dynamics of reconnection and energy release in a flare current sheet. 

If true, then a key result of our work is that flare electron acceleration, as reflected in the observed HXR emission, occurs preferentially when the reconnection is bursty, i.e., when the local increase in ribbon width is especially fast. Furthermore, the particle acceleration occurs during the reconnection outflow.  
Our observations cannot pin down the acceleration mechanism, but they do support a model such as acceleration in magnetic islands, which form in the reconnection outflow and stream down to merge with the arcade of flare loops \citep{drake06a,guidoni16a}. Such a scenario naturally leads to a tight correlation between the UV and HXR emission, since they both arise from the same energy-release process, and to the observed delays between the reconnection onset and the HXR/UV maximum brightness. 

A key question remains, however, as to why the particle acceleration turns on only for an especially fast impulse of reconnection. Furthermore, our observations also show enhancements of the ribbon-front width in some time intervals, particularly between 12:45~UT and 12:48~UT in the early phase, and at some locations (e.g., ROI-2), which do not appear to have any associated HXR signatures. The fast impulse of reconnection therefore appears to be a necessary, but not sufficient, condition for production of non-thermal electrons. The efficiency of particle acceleration may be linked to the proliferation of islands that develop in the coronal current layer and/or to the strength of the reconnection guide field \citep{arnold21a}. A coupled MHD/kinetic model that robustly captures particle acceleration during current-sheet reconnection has been developed recently \citep{arnold21a}. Simulations with this type of capability are now called for, in order to reach the closure with these unprecedented flare-ribbon observations provided by {\em IRIS}. 

A key conclusion 
of our analysis of the ribbon-front structure is the quasi-one-dimensionality of the fronts.
{While the analysis revealed the development of structure along the length of the ribbon with a time cadence  that was linked to hard X-ray emission, overall the ribbons revealed little structure along the direction perpendicular to the ribbon front. 
Such structuring would be expected if energy release in the corona were to spread a substantial distance upstream from a traditional reconnecting current sheet. If magnetic energy release were truly volumetric, energy release would take place at numerous current sheets in a volume and that would be reflected in stacked ribbon fronts in the data. This is not seen, which suggests that while structuring of the ribbon develops along the length of the current sheet and provides evidence for the bursty nature of reconnection, energy release occurs in quasi-two-dimensional flare current sheets.} This observation has important implications for turbulence models of flare energy release and particle acceleration. Certainly traditional models of fully developed turbulence seem inconsistent with the measured structure of the ribbon in the present event. 


\section{Summary}
\label{sec:conclusions}

Using high-resolution, UV imaging observations by {\em IRIS},  
we have analyzed the spatio-temporal evolution of the leading edge, or ribbon front, of the M7.3 SOL2014-04-18T13 flare, and compared with the hard X-ray emissions. We find the following:
\begin{itemize}
\item The ribbon fronts are highly structured in both space and time, exhibiting locally enhanced regions of 1.5-3 Mm in length along the ribbon and 0.6-1.2 Mm in width perpendicular to the ribbon (Figs.~\ref{fig:zebra}, \ref{fig:zebra2796}, and \ref{fig:zebra1400}). This structuring of the ribbon indicates the occurrence of patchy, highly intermittent reconnection in the coronal current sheet. 

\item The evolution of the flare can be divided into two phases. 
      During the early phase, the UV ribbons form with narrow fronts that proceed to widen (Figs.~\ref{fig:zebra2796}e-f and \ref{fig:zebra1400}e-f); the reconnected magnetic-flux change rate rises approximately in lockstep with the front widening (Fig.~\ref{fig:fluxregions}b); and the hard X-rays turn on, but only very gradually at first (Fig.~\ref{fig:overview}).
      
\item During the late phase of the flare, the reconnected magnetic-flux rate drops steeply from its early-time value as the expanding ribbon segments shorten; the widths of the ribbon fronts plateau near their maximum (Fig.~\ref{fig:fluxregions}b); and the UV and HXR light curves rise rapidly to multiple simultaneous peaks (Fig.~\ref{fig:hxr}). 


\item We infer from the data that the evolution of the ribbon-front widths was co-spatial and co-temporal with the UV and HXR emission during the late phase of the flare, but not during the early phase of the flare (Fig.~\ref{fig:tdplot}); this suggests that localized bursts of magnetic reconnection are required for, but do not always lead to, non-thermal electron acceleration. 
\end{itemize}

\acknowledgments

We thank the referee for constructive comments that help improve the clarity of the paper. We thank Vadim Uritsky for insightful discussions. The collaboration leading to these results was facilitated
by the NASA Drive Science Center on Solar Flare Energy
Release (SolFER), Grant No.\ 80NSSC20K0627. S.J.N.\ was also supported by a NASA H-ISFM summer internship at the Goddard Space Flight Center.
J.Q.\ was supported by NASA grants No.\ 80NSSC18K0622 and No.\ 80NSSC19K0269.
C.R.D.\ was supported by NASA's H-ISFM program at GSFC. J.F.D.\ and M.S.\ were also supported by NSF Grants No. PHY1805829 and No. PHY2109083 and NASA Grant No.\ 80NSSC20K1813. J.T.D. was supported by an appointment to the NASA Postdoctoral Program at the NASA Goddard Space Flight Center, administered by Universities Space Research Association under contract with NASA. {\em IRIS} is a NASA small explorer mission developed and operated by LMSAL with mission operations executed at the NASA Ames Research center and major contributions to downlink communications funded by the Norwegian Space Center (NSC, Norway) through an ESA PRODEX contract. {\em SDO} is a mission of NASA's Living With a Star Program. The authors also thank the {\em IRIS} team for help with the data analysis, and the {\em RHESSI} Mission Archive for the data support.


\bibliography{ribbon_all_sim}
\bibliographystyle{aasjournal}



\end{document}